\documentclass[11pt]{article}

\usepackage{amsmath, amssymb, amsthm, amstext, enumerate,hhline}
\usepackage{graphicx, array, fancyhdr, multicol, mathdots, hyperref, stmaryrd}
\usepackage{braket}
\usepackage{fullpage}

\usepackage{authblk}

\renewcommand{\bar}[1]{\overline{#1}}

\newcommand{\poly}{\mathrm{poly}}

\newcommand{\bb}[1]{
	\expandafter\newcommand\expandafter{\csname #1\endcsname}{\mathbb {#1}}
} 
	\bb C

\newcommand{\op}[1]{
\expandafter\DeclareMathOperator\expandafter{\csname #1\endcsname}{#1}
} 
	\op{Enc}
	\op{Dec}

\newcommand{\hads}{\mathcal{H}_{\text{AdS}}}
\newcommand{\hcft}{\mathcal{H}_{\text{CFT}}}

\newcommand{\hamilcft}{H_{\text{CFT}}}
\newcommand{\hamilcftl}{H_{\text{CFT,L}}}
\newcommand{\hamilcftr}{H_{\text{CFT,R}}}

	\theoremstyle{theorem}
		\newtheorem{thm}{Theorem}[section]
		
		\newtheorem{theorem}[thm]{Theorem}

		\newtheorem{claim}[thm]{Claim}
		\newtheorem{conjecture}[thm]{Conjecture}

	\theoremstyle{definition}

		\newtheorem{definition}[thm]{Definition}
	\theoremstyle{remark}

\usepackage{xcolor}

\usepackage{tikz}
\usetikzlibrary{trees}
\usetikzlibrary{calc}

\title{Computational pseudorandomness, the wormhole growth paradox, and constraints on the AdS/CFT duality}
\author[1]{Adam Bouland}
\author[2]{Bill Fefferman}
\author[1]{Umesh Vazirani}
\affil[1]{Dept. of Electrical Engineering \& Computer Sciences, University of California, Berkeley}
\affil[2]{Dept. of Computer Science, University of Chicago}
\date{}

\begin{document}

\clearpage\maketitle

\begin{abstract}

A fundamental issue in the AdS/CFT correspondence is the wormhole growth paradox.
Susskind's conjectured resolution of the paradox was to equate the volume of the wormhole with the circuit complexity of its dual quantum state in the CFT. 
We study the ramifications of this conjecture from a complexity-theoretic perspective. 
Specifically we give evidence for the existence of computationally pseudorandom states in the CFT, and argue that wormhole volume is measureable in a non-physical but computational sense, by amalgamating the experiences of multiple observers in the wormhole. 
In other words the conjecture equates a quantity which is difficult to compute with one which is easy to compute. 
The pseudorandomness argument further implies that this is a necessary feature of any resolution of the wormhole growth paradox, not just of Susskind's Complexity=Volume conjecture. 
As a corollary we conclude that either the AdS/CFT dictionary map must be exponentially complex, or the quantum Extended Church-Turing thesis must be false in quantum gravity.

\end{abstract}

\section{Introduction}

Reconciling general relativity and quantum mechanics into a comprehensive theory of quantum gravity is one of the great unresolved challenges of the last century. The AdS/CFT correspondence \cite{maldacena1999large} has played a central role in guiding progress over the last couple of decades. It conjectures a duality between a theory of quantum gravity in Anti de Sitter space (hyperbolic space) in $d$ dimensions (``the bulk") and a conformal field theory in $d-1$ dimensions (``the boundary").
In particular, states in the bulk are mapped to certain quantum states in the boundary, and operators in the bulk are mapped to certain operators on the boundary. The map between these states/operators is known as the ``AdS/CFT dictionary'' map, or ``dictionary'' for short.

A very interesting puzzle arises in juxtaposing general relativity with the AdS/CFT correspondence. Specifically there are certain solutions to Einstein's equations of gravity known as eternal black holes, which partition space-time into two distinct regions connected by a ``wormhole''. The volume of the wormhole grows steadily for at least an exponential amount of time (in the entropy of the black hole).
Viewed through the AdS/CFT correspondence this poses a conundrum. In the dual space the CFT dynamics of black hole systems are conjectured to be ``scrambling'', i.e. the expectation values of local operators saturate quickly to their equilibrium values, in time that scales nearly linearly in the entropy of the black hole. What quantity in the CFT could possibly then be dual to the wormhole volume \cite{susskind2014original}?

Susskind \cite{susskind2014original,susskind2014addendum,susskind2016computational} argued that the dual quantity is the circuit complexity of the CFT state, defined as the minimum number of local gates required to create that state starting with some reference state\footnote{This particular notion of circuit complexity is uninteresting for classical strings, since the number of gates required to prepare a given $n$ bit classical string can be at most $n$, but is more interesting in the quantum case as preparing a quantum state may require as many as $2^n$ gates.}.
Susskind's intuition was that under time evolution, the circuit complexity of a state should grow at a linear rate until it saturates at its maximum value ($\sim2^n$), which matches the growth of the wormhole volume. 
There has also been a refinement of the conjecture known as ``Complexity=Action''\cite{brown2016complexity,brown2016holographic}, and we refer to these collectively as the ``complexity duality conjectures''.
Much work has been done to test and support these conjectures, both from the gravity \cite{susskind2014addendum,shenker2014multiple,stanford2014complexity,alishahiha2015holographic,lehner2016gravitational,carmi2017comments} and complexity \cite{aaronson2016complexity,bohdanowicz2017universal,jefferson2017circuit,hackl2018circuit,chapman2018toward} perspectives.

In this paper we examine the wormhole growth paradox from a complexity-theoretic viewpoint. We provide evidence that any resolution of the paradox (including the complexity duality conjectures) either violates the quantum Extended Church-Turing thesis (qECT) -- i.e. the belief that all physical processes, including quantum gravity, are efficiently simulable on a quantum computer -- or else requires the AdS/CFT dictionary map to have exponential complexity in certain regimes.

Concretely, it is helpful to start by examining Susskind's Complexity=Volume conjecture. We give evidence that for the allowed quantum states in the CFT, their quantum circuit complexity is exponentially hard to approximate, i.e. it is not ``feelable'' to poly-time experiments.
By contrast we argue that the wormhole volume is ``feelable'' in some general but non-physical sense. 
The duality between a ``feelable'' quantity and an ``unfeelable'' quantity implies that some aspect of this duality must have exponential complexity, as we explain below. While at first sight this might seem to justify the discomfort of complexity theorists with equating computational complexity with a physical quantity, a further examination of our arguments shows that any resolution of the wormhole growth paradox must equate wormhole volume to an ``unfeelable" quantity,
i.e. this discomfort is an inevitable consequence of the paradox.

A key part of this argument appeals to a fundamental notion from cryptography known as computational pseudorandomness, where ensembles of states of low circuit complexity can masquerade as high complexity states to any casual observer. 
While this concept goes back to the early days of complexity-based cryptography \cite{blum1982generateconference,yao1982theory}, it has only very recently been extended to ensembles of quantum states \cite{ji2018pseudorandom,brakerski2019pseudo}.
Unfortunately existing quantum constructions cannot be used in the context of the wormhole growth paradox as only very restricted operations are allowed in the CFT.
Nevertheless, we give evidence for computational pseudorandomness within the domain of the complexity duality conjectures.
Our construction takes its inspiration from symmetric block ciphers such as DES and AES.

Another key part of the paper is the argument that wormhole volume is ``feelable." 
The volume of a wormhole cannot be physically measured by an observer living within the theory, so this is not obvious.
Instead we argue that wormhole volume is (approximately) computable in a more general sense -- by amalgamating the experiences of multiple observers inside the wormhole who cannot physically communicate with one another, or more generally by computing properties of the metric of the semiclassical AdS state. 
While this is non-physical, it suffices to contradict the quantum pseudorandomness properties under two reasonable assumptions: namely that the quantum Extended Church-Turing thesis (qECT) holds in AdS, and that the AdS/CFT dictionary map is of low complexity.
This suggests that Complexity=Volume implies at least one of these assumptions -- the qECT or a low-complexity dictionary -- cannot stand as stated.

A closer examination of the pseudorandomness arguments reveals that it applies more generally to the length of time evolution in the CFT (rather than just circuit complexity). 
This means that any resolution of the wormhole growth paradox implies that either the qECT fails to hold in the AdS, or that the dictionary map must have exponential complexity, particularly when reconstructing the physics of the wormhole interior. 
Our arguments reveal this complexity already kicks in early in the evolution of the wormhole, at times not too much larger than the scrambling time.
See Section \ref{sec:interpretation} for a detailed discussion.

\section{Going deeper into the argument}

 Following prior works \cite{susskind2014original,susskind2014addendum,stanford2014complexity,shenker2014multiple, susskind2016computational}, we model the CFT as an $n$-qubit state where $n$ is the entropy of the black hole. Time evolution is governed by a local Hamiltonian $\hamilcft$. The CFT begins in a fixed reference state $\ket{TFD}$, known as the ``Thermofield Double'' state, which has relatively low energy with respect to $\hamilcft$, and which is dual to a wormhole of zero volume (see \ref{sec:setup} for details). The state evolves according to $\hamilcft$, with the evolution perturbed by inserting a small number $\ell$ of one-qubit operators -- with $\ell \ll n$ -- called ``shocks''. That is consider states of the form
\begin{equation}
\label{eq:shockstate}
    e^{-i\hamilcft t_\ell} \mathcal{O}_\ell e^{-i\hamilcft t_{\ell-1}} \mathcal{O}_{\ell-1} \ldots \mathcal{O}_1 e^{-i\hamilcft t_0} \ket{TFD} .
\end{equation}

Note that each shock increases the energy of the system by an $O(1)$ term and perturbs the geometry of the wormhole. The limitation on the number of shocks ensures that the total energy added to the system is much less than the black hole's size, and the geometry of the wormhole is roughly preserved.
The Complexity=Volume conjecture states that the circuit complexity of such states is equal to the wormhole volume of their dual states (see section \ref{sec:formalc=v}). 
Indeed \cite{shenker2014multiple,stanford2014complexity} gave supporting evidence for these conjectures based on quantum gravity calculations. 

\subsection{Pseudorandomness in the CFT}

We first argue that the complexity, or more generally the length of time evolution, of states of the form (\ref{eq:shockstate}) is not ``feelable'', in the sense that even given a small number $k$ copies of the state, no efficient algorithm can approximate the time $t = \sum_{j=0}^{\ell} t_j$ for which it evolved\footnote{This can be seen as a proxy for the circuit complexity of the state relative to $\ket{TFD}$}. 
To do so we will argue that ensembles of such states are ``pseudorandom'' -- that is they are effectively indistinguishable from Haar-random states, and hence from one another as well.
That is, if one only had access to $k$ copies of states of the form (\ref{eq:shockstate}), and did not know the pattern of shocks applied, then it would be very difficult to distinguish them from Haar-random states. 

It turns out that information-theoretic techniques will not work to establish this conclusion. 
The standard approach would be to show that such states are $k$-designs over the choice of random shock patterns -- that is they match the first $k$ moments of the Haar measure -- which would imply that $k$ copies of a random such state are information theoretically indistinguishable from $k$ copies of a Haar random state. 
However the total number of allowable CFT states is too small to even accommodate a $1$-design\footnote{In particular, a $1$-design on $n$ qubits must contain at least $\sim 2^n$ distinct states \cite{roberts2017chaos}.} \cite{roberts2017chaos} due to the limited number $\ell \ll n$ of shocks allowed in the CFT.

Instead we will use the notion of \emph{computational} pseudorandomness.
The basic idea is that one can efficiently create ensembles of quantum states which are information theoretically distinguishable from random, but any measurement which distinguishes them must have exponential complexity~\cite{ji2018pseudorandom}. We will describe this notion in more detail in Section \ref{sec:pseudorandomnessintro}. 
Unfortunately the allowable CFT states do not support the prior pseudorandom construction of~\cite{ji2018pseudorandom} based on post-quantum cryptography.
Instead we create a quite different construction tailored to the constraints on the CFT states.

A key property of the CFT that we will use in our construction is that black holes are highly ``scrambling'' -- that is after some short time $t_{scr}$, evolution by $e^{-i\hamilcft t_{scr}}$ diffuses local perturbations throughout the system (see e.g. \cite{harrow2019separation}).
One might be tempted to confuse this scrambling property with pseudorandomness, but of course evolution by $\hamilcft$ itself is not pseudorandom, because if one knows $\hamilcft$ then one can time evolve backwards to determine the length of time evolution of a state of the form $e^{-i\hamilcft t}\ket{TFD}$.
Therefore to achieve pseudorandomness, we will alternate this ``scrambling'' unitary $U = e^{-i \hamilcft t_{scr}}$ with ``shocks''. Consider states of the form
\begin{equation}
    \ket{\phi_{\ell}} = U \mathcal{O}_\ell U  \mathcal{O}_{\ell-1} \ldots \mathcal{O}_1 U \ket{TFD} .
\end{equation}
Where each $\mathcal{O}_i$ is a random one-qubit Pauli operator. The basic idea is that each ``shock'' induces a branching point in the evolution -- so if there are $k$ shocks applied then there are $\sim 2^k$ possible states one could be in after $k$ iterations. 
Intuitively determining if a given state $\ket{\psi}$ was one of these $\sim 2^k$ states is a difficult search problem -- which makes it difficult to distinguish such states from random. 
We conjecture that these states are indeed computationally pseudorandom.

To provide evidence for this conjecture, we consider a simplified black-box setting where $U=e^{-i \hamilcft t_{scr}}$ is modeled by a random black-box unitary, and complexity is measured relative to $U$ (i.e. $U$ is assumed to have unit cost).
Here $U$ is fixed and accessed via black box quantum queries to $U$ and $U^{\dagger}$, and the shocks $O_j$ are unknown.
We argue that the state $\ket{\phi_{\ell}}$ is pseudorandom in this model, because the number of black box queries for distinguishing $\ket{\phi_{\ell}}$ from a Haar random state scales as $2^{\ell/2}$. 
We prove this lower bound rigorously in a toy model of the system where $U$ is a permutation of computational basis states, and sketch the proof in the quantum case as well.
We conjecture, assuming that the evolution of the black hole is sufficiently ``scrambling,'' that some version of this black-box security property holds even when $U$ is replaced by $e^{-i\hamilcft t_{scr}}$.

The black box abstraction above models the computational aspects of the question, but does not model certain physical aspects. 
For instance, the states (\ref{eq:shockstate}) are low energy states of the CFT, and therefore energy measurements leak information about when/where the shocks are applied, weakening the security of our construction. 
This requires careful consideration (see Section \ref{sec:energyissues}), but we argue that our construction remains secure against energy measurement attacks.
The key issue is that the number of copies of the state required to break our arguments via energy measurements is far larger than the number of copies of the state needed to approximate the volume of the AdS wormhole.
Therefore we conjecture that a sufficiently strong form of pseudorandomness exists in the CFT to imply the conclusions that we outline below.

\subsection{Consequences for AdS/CFT}

We have provided evidence for pseudorandomness in the CFT.
In Section \ref{sec:volumeeasy}, we also argue that wormhole volume is ``feelable'', but not in the sense that there exist physical experiments that observers could perform in the wormhole to estimate the volume. 
Instead, we argue that the wormhole volume is an efficiently computable property of the metric of the semi-classical AdS states we consider in our construction -- and therefore should be easy to compute if one can extract coarse properties of the metric from the AdS state (which we emphasize is a \emph{classical} property of the AdS state).
For instance, such properties of the metric could be extracted in a model where one can post-process the simulated experiences of several observers in the bulk, who may be space-like separated until they encounter the singularity. 
Nevertheless, this is sufficient for a contradiction as argued below.

For clarity, we outline the argument assuming Complexity=Volume, and we note that similar arguments suggest that no resolution of the wormhole growth paradox can escape these consequences, barring a fundamental change to the structure of the AdS/CFT correspondence (See Sections \ref{sec:generalarg}, \ref{sec:interpretation}). 
Suppose that (A) the dictionary map is computable in quantum polynomial time, and (B) the quantum Extended Church-Turing thesis holds for quantum gravity -- 
i.e. for a suitable discretization of quantum gravity, the evolution of a state in the bulk is efficiently simulable by quantum computers.
Consider picking two pseudorandom CFT states $\ket{\psi_1},\ket{\psi_2}$ with vastly different (e.g. $n^2$ vs. $n^3$) complexities.
By the pseudorandomness property, they should be computationally indistinguishable from random, and hence from each other as well.
Now imagine passing these states through the dictionary map $\Phi$.
By Complexity=Volume, these correspond to vastly different wormhole volumes.
These different volumes can be differentiated by postprocessing the experiences of several non-communicating observers in the AdS space.

The key point is that, if quantum gravity is efficiently simulable on a quantum computer (i.e. the qECT holds), one can not only efficiently simulate the physics of the AdS system, but one can also combine the experiences of multiple observers in multiple instantiations of the AdS system, and postprocess their observed outcomes. 
Composing these steps -- application of the dictionary, simulating the AdS experiments using a quantum computer, and classical postprocessing -- one obtains an efficient quantum circuit distinguishing $\ket{\psi_1}$ and $\ket{\psi_2}$. This contradicts the pseudorandomness property, suggesting that statements (A) and (B) cannot be simultaneously true.

\section{Computational pseudorandomness through the quantum lens}

\label{sec:pseudorandomnessintro}

Modern cryptography starts with a switch from \emph{information theoretic} notions of security to \emph{computational} notions of security, leading to a rich collection of cryptographic primitives -- such as public-key cryptography, pseudorandomness, zero-knowledge proofs for NP, homomorphic encryption, etc. -- all of which are believed to be impossible in the information theoretic setting. One of the earliest such applications was pseudorandomness \cite{blum1982generateconference,yao1982theory,goldreich1986construct}. 
The basic idea is that one can construct a deterministic procedure that maps a short uniform seed to a much longer bitstring. While the output bits are no longer uniformly distributed, the resulting distribution is ``pseudorandom'', or indistinguishable from uniform by any computationally bounded adversary.

Recently, Ji, Liu and Song \cite{ji2018pseudorandom}
defined the notion of a pseudorandom quantum state (PRS) ensemble, which is a quantum adaptation of these ideas.
They define an ensemble of quantum states on $n$ qubits, denoted $\lbrace \ket{\Psi_k} \rbrace$ indexed by a key, $k\in\{0,1\}^{poly(n)}$, to be a PRS if the following hold:
\begin{itemize}
    \item There is a polynomial-time quantum algorithm that, when given as input the key, $k$, generates the state $\ket{\Psi_k}$.
    \item For any polynomial $p(n)$, the state $\mathbb{E}_k \left(\ket{\Psi_k}\bra{\Psi_k}^{\otimes p(n)}\right)$ cannot be distinguished from \newline $\mathbb{E}_{\ket{\phi}\sim \mathcal{H}} \left(\ket{\phi}\bra{\phi}^{\otimes p(n)}\right)$ by any polynomial-time quantum algorithm with nonnegligible probability, where $\mathcal{H}$ denotes the Haar measure on $\left(\mathbb{C}^2\right)^{\otimes n}$.
\end{itemize}
In other words, a PRS is an ensemble of efficiently constructible quantum states which are indistinguishable from Haar-random quantum states by any efficient quantum algorithm. 
They remain indistinguishable even if one is allowed access to arbitrarily polynomially many copies of the state\footnote{We note that given arbitrarily polynomially many copies of a state, such states are \emph{always} information theoretically distinguishable from random \cite{brandao2016efficient}, so this notion is fundamentally computational.}.

To see how this is possible, let us briefly recap Ji, Liu and Song's construction \cite{ji2018pseudorandom}. Consider states of the form:
\begin{equation} \label{eq:randomsuperpositionstate}
\ket{\psi_r}=\frac{1}{2^{n/2}} \sum_{x\in \{0,1\}^n} (-1)^{r(x)} \ket{x}\end{equation}
where $r(x)$ is a truly random function of $x$.
They show that such states are information theoretically indistinguishable\footnote{Ji, Liu and Song showed this is true for random phases over higher roots of unity, and this was recently extended to pseudorandom binary phases by Brakerski and Shmueli \cite{brakerski2019pseudo}. } from Haar-random states \cite{ji2018pseudorandom,brakerski2019pseudo}. 
However, they do not form a PRS, because they are not themselves efficiently generable -- as there are $2^{2^n}$ random functions $r$ a simple counting argument says most $\ket{\psi_r}$ cannot have poly-sized quantum circuits.
Now the key insight is to replace $r$ with a pseudorandom function $f_k$, depending on a secret key $k$ of polynomial length, which looks random\footnote{The pseudorandomness property states that $f$ is indistinguishable from a random function $r$ with high probability over the random choice of the key $k$.} to any poly-time quantum algorithm, but is in fact efficiently computable.
Explicit constructions of such functions $f_k$ are possible assuming there exists a one way function which is secure against quantum computers \cite{zhandry2012construct}.
Consider states of form:
\[\ket{\psi_k}=\frac{1}{2^{n/2}} \sum_{x\in \{0,1\}^n} (-1)^{f_k(x)} \ket{x}\]
Where, for a given pseudorandom function $f_k$, we take the ensemble of states indexed by keys $k$. This ensemble is efficiently constructible by definition -- given $k$, one simply needs to compute $f_k$ in superposition (which can be done with one query to $f_k$, using the ``phase-kickback'' trick) -- and are computationally indistinguishable to states of the form (\ref{eq:randomsuperpositionstate}).
Therefore by the transitivity of indistinguishability, these states are computationally indistinguishable from Haar-random states.
In short, so long as pseudorandom functions secure against quantum adversaries exist (and hence, as long as post-quantum cryptography is possible), one can efficiently prepare quantum states which look Haar-random to any efficient quantum algorithm even with access to many copies of the state.

The existence of PRS's has deep implications for quantum circuit complexity. In particular, we can use this to conclude that it is very difficult to estimate the quantum circuit complexity of a black-box quantum state, even given many copies of the state.
This follows because Haar random states have exponential circuit complexity with high probability (due to a simple counting argument), yet are indistinguishable from PRS states with low circuit complexity -- therefore an algorithm that distinguishes high vs. low quantum circuit complexity could break the PRS security guarantee.
Therefore the quantum circuit complexity of a quantum state is very difficult to compute, even approximately.

This observation is an important conceptual starting point for our argument.
In our argument, we will construct ensembles of PRS states within the domain of validity of the complexity duality conjectures -- and we will use the fact that it is very difficult to estimate the complexities of these states. 
It turns out that Ji, Liu, and Song's construction will not work directly here, because these states will not be compatible with various constraints associated with AdS/CFT \footnote{For instance, these states are very entangled, and do not obey the Ryu-Takayanagi formula \cite{ryu2006holographic,ryu2006aspects}}.
Therefore one of the technical contributions of our work will be to construct a \emph{new ensemble of pseudorandom quantum states} which is designed to fit the constraints of the AdS/CFT setup.
Interestingly, and unlike the Ji \emph{et al.} result, our result does not follow from a reduction to a classical pseudorandom primitive, and instead we prove pseudorandomness in a black-box setting.
Our construction will be inspired by symmetric block ciphers such as DES and AES, which we will discuss in more detail shortly.

\section{Our results}

We now describe the relevance of pseudorandom states to the wormhole growth paradox and the AdS/CFT correspondence. 

\subsection{Setup}
\label{sec:setup}

We begin with a toy model of the AdS/CFT correspondence, which is general enough to capture prior models of the correspondence. We will assume that the CFT has an associated Hilbert space $\hcft$, and that the AdS space has an associated Hilbert space\footnote{In the spirit of complementarity \cite{susskind1993stretched}, one could also consider a model in which there is not a single Hilbert space for the entire AdS space, but rather only a Hilbert space for each causal patch $P$ of the AdS space. Then there would be a family of dictionary maps $\Phi_P$, one for each causal patch $P$ of the AdS space, which maps from $\hcft$ to the Hilbert space of that causal patch $\mathcal{H}_{\text{AdS,P}}$. We note our argument still applies to this generalization -- it says that the assumptions (A) the qECT and (B) \emph{all} of these maps $\Phi_P$ are of low complexity, are inconsistent.} $\hads$. 
Following \cite{pastawski2015holographic}, for simplicity we will assume that both $\hcft$ and $\hads$ are described by systems of $n$ qubits (and $n'$ qubits, respectively) for some sufficiently large $n,n'=\text{poly}(n)$ -- this is to ensure that the circuit complexity of states in $\hcft$ is well-defined, and to allow us to prove certain facts about how the complexity of certain objects scale with $n$.

We further assert that there is some subset $S\subseteq \hcft$ and $T\subseteq \hads$ to which the AdS/CFT correspondence applies\footnote{For instance $T$ could represent the subset of states which correspond to smooth semiclassical gravity and $S$ could represent those states which obey the Ryu-Takayanagi formula.}.
Within these subsets there exists a state dictionary map $\Phi:S\rightarrow T$ which maps CFT states to their AdS states, and an operator dictionary map $\Phi_{\text{op}}$ which maps operators on $S$ to operators on $T$.
(We note that $\Phi$ and $\Phi_{\text{op}}$ are closely related to one another by switching from the Heisenberg to Schr\"{o}dinger pictures, so we will often just denote this $\Phi$ when the context is clear). 
This model is general enough to capture prior toy models of the correspondence \cite{pastawski2015holographic,kohler2018complete}.
Our only additional assumption will be that there exists local time evolution Hamiltonians $\hamilcft$ and $H_{\text{AdS}}$ which control the time evolutions of the systems, so that $H_{\text{AdS}} = \Phi_{\text{op}}\left(H_\text{CFT}\right)$ \footnote{However we need not assume that $\Phi$ preserves the locality of all Hamiltonians as in \cite{kohler2018complete}.}.
We note that this assumption on the AdS side implies the qECT for the AdS space -- as one can efficiently simulate local Hamiltonian evolution with standard quantum computers. For an interpretation of our results without this assumption see Section \ref{sec:interpretation}.

\subsection{Formal statement of C=V}
\label{sec:formalc=v}

The Complexity=Volume Conjecture (C=V) is stated in the context of the two-sided AdS black hole, a solution to the classical equations of GR. 
In this particular system the CFT is divided into two sections, denoted left and right, each with their own Hamiltonian $\hamilcftl$ and $\hamilcftr$, so $\hamilcft = \hamilcftl+\hamilcftr$. 
The conjectured dual of this black hole is the Thermofield Double State which up to normalization is 
\[\ket{TFD}=\sum_{i} e^{-E_i/\beta} \ket{i}_L \ket{i}_R\]
Where the $\ket{i}_L$ ($\ket{i}_R$) denotes the energy eigenstates of $\hamilcftl$ ($\hamilcftr$, respectively), $E_i$ denotes the energy of these eigenstates, and $\beta$ is an inverse temperature parameter. 
The state of this system after time $t$ under its natural time evolution is given by
\[\ket{TFD(t)} = e^{-i\hamilcft t} \ket{TFD}.\]
While it is not essential to the argument, following Stanford and Susskind \cite{stanford2014complexity} we assume $n=S$, i.e. the number of qubits is the entropy of the two-sided black hole we are modeling.

Define the relative circuit\footnote{Note circuit complexity is not entirely well-defined on CFT states \cite{brown2016complexity}, as CFT states are not over tensor products of qubits, and there is not a canonical choice of universal gate set. While there have been attempts to define a continuum version of complexity which applies to QFT and CFT states, e.g. \cite{jefferson2017circuit,hackl2018circuit,chapman2018toward}, here we will consider a qubit-based definition of circuit complexity as a toy model of CFT complexity.} complexity $\mathcal{C}_\epsilon(\ket{\psi},\ket{\phi})$ to be the minimum number of two-qubit gates in a quantum circuit $U$ such that $||U\ket{\psi} - \ket{\phi}||\leq \epsilon$ in some distance metric. 
It is known that the volume of the wormhole $V(t)$ grows linearly with time in classical GR.
The Complexity=Volume conjecture of Susskind \cite{susskind2014original,susskind2016computational} loosely states the following: that for some suitable choice of $\epsilon$ and constants $c,c'$, and for times $0\leq t\leq O(2^n)$, that 
\[\mathcal{C}_\epsilon \left(\ket{TFD},\ket{TFD(t)}\right) \approx c\cdot V(\Phi(\ket{TFD(t)})) \approx c' t.\]

In other words, the complexity of the state of the CFT grows linearly with time (at least for an exponential amount of time), just as a wormhole volume behaves in classical GR. 
It is precisely this statement -- that the complexity of a time-independent Hamiltonian evolution should grow linearly with time -- that Aaronson and Susskind \cite{aaronson2016complexity} and Bohdanowicz and Brandao \cite{bohdanowicz2017universal} have made progress towards proving.

In some sense the equation stated above is relatively superficial: it simply states that the Complexity and the Volume both have the same behavior for this particular state under this particular Hamiltonian evolution. 
This correlation does not establish a causal relationship between the Complexity and Volume - it could simply be a coincidence for this special case. 
Therefore, the Complexity=Volume goes further than the above statement, and applies to a wider domain of quantum states.

In particular, let us define the family of ``small perturbations'' of the normal time evolution of this system.
We will take our inspiration from prior works \cite{shenker2014black, shenker2014multiple, susskind2014addendum, stanford2014complexity} which defined precisely such a model.
In these works, the authors considered perturbing the evolution of the system by one or more low-energy local CFT operators $\mathcal{O}$; in the physics language these operators generate ``shock waves'' in the system (since they are inserted at an precise/infinitesimal time) by ``tapping'' (i.e. perturbing) the boundary CFT state.
They argued that the complexity of the resulting time evolutions matches the behavior of the volume of the wormholes under these perturbations, building evidence for the C=V conjecture. 
We note that they considered the behavior only for ``low energy'' shocks -- in other words the energy of the state before and after the application of the operator is increased by only $O(1)$ thermal quanta of the black hole. 
The purpose of this restriction is to ensure the shock wave is small enough such that the wormhole geometry is still well defined after this perturbation, and to allow for simple perturbative calculations of the volume in this regime.

In the spirit of this prior work, let us formally define a qubit model of these states.

\begin{definition}
\label{def:setofstatesFK}
Let $F_{L,T} \subseteq \hcft$ denote the subset of boundary states that can be generated by perturbing the time evolution of the TFD state up to time $T$ with any $\ell \leq L$ local gate insertions. This includes all states of the form:

\[
\ket{\psi} = e^{-i\hamilcft t_{\ell}} \mathcal{O}_\ell e^{-i\hamilcft t_{\ell-1}} \mathcal{O}_{\ell-1} \ldots \mathcal{O}_1 e^{-i\hamilcft t_0} \ket{TFD}
\]
where each $\mathcal{O}_i$ is a local quantum gate, and $\sum_i t_i \leq T$. 
\end{definition}

We note that applying a local quantum gate will perturb the energy of the system by a constant (since we have assumed that the Hamiltonian $\hamilcft$ is local in the boundary), and therefore this definition closely models the CFT states considered in prior work.

With this definition in mind, we can formally state the C=V conjecture:
\begin{conjecture}[Complexity=Volume Conjecture \cite{susskind2014original,susskind2016computational}]
For any $0\leq T \leq O(2^n)$, then there exists an $L>n^\delta$ for some $\delta>0$, an $\epsilon>0$ and a constant $c$ such that 
for any state $\ket{\psi}$ in $F_{L,T}$, 
\[\mathcal{C}_\epsilon \left(\ket{TFD},\ket{\psi}\right) \approx c \cdot V(\Phi(\ket{\psi}))\]
\end{conjecture}

That is, C=V conjectures a more causal relationship between the Complexity and Volume, which holds for a larger family of states -- so it makes sense to say that they are dual to one another. 
This family of states includes both the normal time evolution of the two sided AdS black hole, as well as reasonable perturbations of it -- for instance perturbations with ensure the AdS still has a wormhole and has a reasonably smooth geometry. (This is clearly necessary, to ensure the volume side of the equation is well defined). 
While the explicit value of $L$ was not defined in prior works, we take $L>n^\delta$ to allow for a significant amount of perturbation to the initial state, but still far less than the amount required to destroy the wormhole geometry (say $L\approx n$). 
This exact idea was sanity-checked by Stanford and Susskind \cite{stanford2014complexity} and is one of the key pieces of evidence for the conjecture.

\subsection{The core argument}
\label{sec:generalarg}

We now state the core of our argument, which applies not only to the complexity duality conjectures, but to any resolution of the wormhole growth paradox.

Consider any resolution of the wormhole growth paradox, in which some quantity $Q(\ket{\psi_{CFT}})$ is dual to the volume of the corresponding AdS wormhole, which is a function of the CFT state only. (For instance, Susskind's conjecture is $Q(\ket{\psi_{CFT}}) = \mathcal{C}_\epsilon \left(\ket{TFD},\ket{\psi_{CFT}}\right)$).
In other words, we have that for any $0\leq T \leq O(2^n)$, then there exists an $L>n^\delta$, an $\epsilon>0$ and a constant $c$ such that for any state $\ket{\psi}$ in $F_{L,T}$, then
\[Q \left(\ket{\psi}\right) \approx c \cdot V(\Phi(\ket{\psi}))\]
Since the wormhole volume grows linearly with time (up to small corrections due to the shocks \cite{shenker2014multiple}), then $Q$ must grow linearly with time as well.

Suppose that we can construct a PRS ensemble $\{\ket{\Psi_k}\}$ in the CFT, in which each state in the ensemble, $\ket{\Psi_k}$, is in $F_{L,T}$ -- that is, the domain of validity of the wormhole growth paradox. 
More concretely, suppose that with $c(n)=O(\text{poly}(n))$ copies of the state $\ket{\Psi_k}$ for a random $k$, distinguishing it from random requires exponential time for a quantum computer. 

Furthermore, suppose that for each such state, the evolution time of the CFT states $\ket{\Psi_k}$ is say $O(n^2)$.
By our assumed resolution of the wormhole growth paradox, $Q(\Psi_k)$ should be $O(n^2)$, as well as the volume of the dual wormhole.
Now consider a second ensemble of pseudorandom states $\{\ket{\Psi'_{k'}}\}$ with the same properties but with Q value and wormhole volume\footnote{We note that our arguments would also work with smaller sized wormholes -- see Section \ref{sec:volumeeasy}.} say $O(n^3)$.
We will argue for the existence of such states in detail in Section \ref{sec:customPRS}.

Since these ensembles of states are indistinguishable from random, in particular they are indistinguishable from one another as well\footnote{This follows because if they were distinguishable from one another with bias $b$, by considering a hybrid argument with a Haar-random state as the hybrid, then one of them must be distinguishable from random with bias $\geq b/2$.}.
That is, no quantum algorithm can distinguish between any polynomially bounded number, $m$, copies of a state $\ket{\Psi}$ drawn either from the first PRS ensemble with $Q=O(n^2)$ or from the second with $Q=O(n^3)$
without using computation time $T(n)$ which is exponential.

Now the interesting point is that these ensembles of states are not easily distinguishable from one another in the boundary, by construction. 
However, they have substantially different wormhole volumes. We now claim that the wormhole volumes of these AdS states \emph{can} be efficiently distinguished. Suppose that with $c(n)$ copies of the AdS states $\Phi(\ket{\Psi_k})$ or $\Phi(\ket{\Psi'_{k'}})$ that one could roughly estimate\footnote{Say to constant multiplicative error.} the length of the wormholes in those states in time $t(n)\leq \poly(n)$ using some algorithm $\mathcal{A}$ (See Section \ref{sec:volumeeasy}). 
Here we are assuming the quantum Extended Church-Turing thesis -- so that our distinguishing experiments in the bulk are efficiently simulable on a quantum computer.

Then by composing $\mathcal{A}$ with the dictionary map $\Phi$, one could distinguish the $\Psi_k$ vs $\Psi'_{k'}$ in the CFT -- as these have grossly different wormhole volumes. 
Therefore, if $\mathcal{C}_\epsilon(\Phi)$ denotes the gate complexity of $\epsilon$-approximately implementing $\Phi$ on a quantum computer, then we immediately have that
\[\mathcal{C}_\epsilon(\Phi) t(n) \geq T(n) \]
And so,
\[\mathcal{C}_\epsilon(\Phi) \geq T(n)/t(n) \geq \text{exp}(n)\]
For some exponential function $\text{exp}(n)$.
In other words, if it is relatively easy to approximate the volume of a wormhole in the AdS Hilbert space (at least to accuracy good enough to distinguish macroscopically different wormhole volumes), and the quantum Extended Church-Turing thesis is true, the dictionary must be very complex -- as the pseudorandomness argument says that distinguishing these states on the CFT side must be very difficult.

\section{Pseudorandomness in the CFT}
\label{sec:customPRS}

As we have seen in the previous section, the key step in applying pseudorandomness to constrain AdS/CFT in light of the wormhole growth paradox is to give evidence that there exist families of PRS's within the conjecture's domain of validity $F_{L,T}$ for small $L$ -- i.e. reasonable perturbations of the underlying dynamics. 
In this section, we will provide such a construction and argue for its security in a certain restricted black-box setting. 

Recall that $F_{L,T}$ is the subset of $\hcft$ reachable by starting from the TFD state and perturbing the normal time evolution of the system for time up to $T$ with up to $L$ local perturbations (see Definition \ref{def:setofstatesFK}). 
In this argument we will only consider evolutions for polynomial amounts of time, i.e. $T\leq poly(n)$.
Consider the time evolution under $\hamilcft$ of the initial CFT state $\ket{TFD}$ on $n$ qubits. 
Let $t_{scr}$ denote the scrambling time of the system, i.e. the time at which $e^{-i\hamilcft t_{scr}}$ behaves as a two design and delocalizes local operators. 
It is usually assumed that $t_{scr}=O(n\log n)$ by the fast scrambling conjecture \cite{sekino2008fast}.

The basic idea of the pseudorandomness construction is the following: clearly the normal time evolution of the TFD state cannot be a PRS ensemble, since one could always evolve the state backward using $\hamilcft$ and test if one arrives at the TFD state.
But now suppose that we insert $\ell$ random local operators (say random one qubit Paulis on the first qubit) into this evolution, where the insertions are separated in time by many multiples\footnote{Note that even a constant multiple of the scrambling time would suffice here, which would allow one to instantiate our arguments with minimally sized wormholes.} $m$ of $t_{scr}$. 
This modification hinders the backwards time evolution distinguishing algorithm, because applying a Pauli operator to a generic state renders it nearly orthogonal to itself -- and therefore each of the $4^\ell$ possible operator insertion patterns results in an (essentially) distinct state. 
So intuitively it should be difficult to distinguish this large basis of states from a truly random state using the forward and backward time evolution of the system.
In fact, we can make this intuition rigorous, and show that there is no algorithm to ``discover the Pauli pattern'' of a given state of this form. This then allows us to use this pattern as the secret key of a PRS ensemble.

More formally, consider evolving the state for $\ell$ discrete epochs of $m\times t_{scr}$ (for total time $\ell\times m\times t_{scr}$). 
Here we think of $\ell$ as being relatively small (say $\ell=n^{\delta}$ for some small $\delta>0$) while $m=O(1)$ is a constant greater than or equal to 2. (One could also consider larger $m$).
Now consider inserting, between each evolution $i$ for time $m\times t_{scr},$ a randomly chosen one qubit Pauli gate $\mathcal{O}_i$ on the first qubit into equally spaced points of this time evolution. That is, if we let $U=e^{-i\hamilcft mt_{scr}}$, then our final state for $\ell=2$ might look like
\[UY_1UZ_1U\ldots \ket{TFD}\]
where the total number of insertions of Pauli operators is $\ell$, the operator $U$ is applied $\ell$ times.
We will take this to be our PRS ensemble.
\begin{definition}
Given a unitary $U$, let the distribution $D_{U,\ell,\ket{\phi}}$ (or, when $\ell$ and $\ket{\phi}$ are implicit, simply $D_U$) over quantum states be defined as follows: begin with the state $\ket{\phi}$, and alternatively apply $U$ and random, independently chosen Paulis on the first qubit $\ell$ times. If the pattern of Paulis chosen is $k\in\{I,X,Y,Z\}^\ell$, then call the resulting state $\ket{\Psi_{k}}$. 
\end{definition}

Here the secret key of the PRS ensemble is the Pauli insertion pattern $k\in\{I,X,Y,Z\}^\ell$. 
By construction, by choosing $U=e^{-i\hamilcft mt_{scr}}$ and $\ket{\phi}=\ket{TFD}$, the states $\ket{\Psi_k} \in \text{supp}(D_U)$ lie within $F_{\ell,mt_{scr}\ell}$.
.

We note that the volume of the wormholes of these states will be approximately $O(\ell t_{scr})$.
To see this, note that each operator insertion (an $O(1)$ perturbation to the energy of the system) should only perturb the wormhole length by at most a factor of $t_{scr}$ due to possible ``switchback'' cancellations -- this was precisely the calculation made by Stanford and Susskind \cite{stanford2014complexity}. 
On the other hand, normal time evolution by time $t$ increases the wormhole volume by roughly a factor of $t$. Therefore one can see that the volume of the AdS wormhole for these states should be $O(m\ell t_{scr}) \pm O(\ell t_{scr})$, which is bounded between $O((m-1)\ell t_{scr})$ and $O(m\ell t_{scr})$.
As we choose $\ell=O(n^{\delta})<< n$, and $m=O(1) \geq 2$, to lowest order in $n$ the wormhole volume of all states in the ensemble is $O(\ell t_{scr})$.
In other words, our perturbations to the time evolution have not affected the wormhole volumes much.

We now claim that these states are very difficult to distinguish from Haar-random -- and from one another as well.
More formally, we claim that

\begin{conjecture}
Let $U=e^{-i\hamilcft mt_{scr}}$ for $m\geq 2$.
Given $c(n)$ copies of a state $\ket{\psi}$, which is either
\begin{itemize}
    \item Haar-random, or else
    \item drawn from $D_U$
\end{itemize}

Then no quantum algorithm can distinguish which is the case with nonnegligible bias in time less than $2^{\Omega(\ell)}$, even given the description of the local Hamiltonian $\hamilcft$.
\label{conj:pseudorandomness}
\end{conjecture}

Here $c(n)$ is the number of copies of the state that we need to estimate the volume of the wormhole (see Section \ref{sec:volumeeasy}). As our arguments will only ever invoke $c(n)$ copies of the state, there is no need to conjecture the security of this construction against arbitrary polynomially many copies of the state. We will argue that $c(n)$ is relatively small. Furthermore, due to certain physical considerations, we would not expect security to hold against arbitrary polynomially many copies ($\gg c(n)$), because measuring the energy of the system leaks information which weakens the security of the construction. See Section \ref{sec:energyissues} for details.

In order to build evidence for this conjecture, we will sketch a proof of this security property in the black box setting:
\begin{claim}
Given a black-box unitary $U$ and an arbitrary $\text{poly}(n)$ copies of a state $\ket{\psi}$, which is either
\begin{itemize}
    \item Haar-random, or else
    \item drawn from $D_U$
\end{itemize}
Then no quantum algorithm can distinguish which is the case with nonnegligible bias in time less than $2^{\Omega(k)}$, even given the ability to apply $U$ and $U^{-1}$ as a black box.
\label{theorem:quantumpseudorandomness}
\end{claim}

In other words, we will sketch a proof of our conjecture in a black-box model, where evolution for a scrambling time is modeled by a black-box unitary. In fact our black-box security proof is slightly stronger than what we need for our conjecture, as security of the black-box model holds against arbitrary polynomially many copies of the state.
We are therefore conjecturing that time evolution by a scrambling time is sufficiently unstructured that it acts like a generic unitary.

Before moving on to proving these statements, let us briefly comment on the analogies between our construction and prior constructions in classical cryptography. At a high level, this pseudorandomness construction can be seen as a version of a symmetric block cipher.
Many block ciphers such as AES and DES perform encryption by alternating applying a random-looking function (say a pseudorandom function with some secret key) and permuting the bits of the string in some pre-determined manner (see e.g. \cite{luby1988construct}).
Our quantum PRS construction has precisely this form: we alternate between applying Paulis according to a secret key and a known random unitary.
The key difference between our construction and prior classical cryptosystems is that the random function we apply between permutations is relatively weak -- it is merely a single random Pauli -- rather than a full-fledged PRF as in e.g. DES \cite{luby1988construct} -- due to the constraints of the C=V setup.
Nevertheless we show that the complexity of breaking this cryptosystem in a black-box manner is high due to the fact that we iterate these operations many times, boosting the difficulty of breaking the construction (as has been argued for classical block ciphers as well \cite{aiello1998security}).

We now sketch a proof of our main pseuodorandomness claim (Claim \ref{theorem:quantumpseudorandomness}.)
Before we do so, we will first introduce a toy classical model of this process, for which we will rigorously prove black-box pseudorandomness under quantum queries, in section \ref{sec:toypseudorandomness}. 
This section will be particularly simple, and yet captures all the essential features of our main proof.
We will then sketch a proof of its generalization to our quantum setting (Claim \ref{theorem:quantumpseudorandomness}) in Section \ref{sec:quantumpseudorandomness}, and discuss further physical considerations for this model in Section \ref{sec:energyissues}.

\subsection{Pseudorandomness for a toy classical model}
\label{sec:toypseudorandomness}

In this section we will describe a toy classical model inspired by our candidate PRS construction, and prove a black box pseudorandomness property. We will sketch how to generalize these results to the quantum setting in the next section.

Consider a uniformly random but fixed permutation $\sigma \in S_{2^n}$ which acts on $n$-bit strings $x\in \{0,1\}^n$. Also consider the operator $X_1$ which flips the first bit of the string, and let $\pi = \sigma \circ X_1$ -- i.e. $\pi$ is obtained by flipping the first bit of the string and then applying $\sigma$. 

Now let $D_\sigma$ be the distribution over $n$-bit strings generated by the following process: start with the input string $0^n$, then apply a uniformly random sequence of $\ell$ $\sigma$'s and $\pi$'s -- i.e., at each of the $\ell$ steps we apply either $\sigma$ or $\pi$, each with probability $1/2$ -- and then output the resulting $n$ bit string.
Clearly, when $\ell$ is very large ($k \gg n$) then this distribution converges to uniform in the case that $\sigma$ was chosen uniformly from $S_{2^n}$. On the other hand, if $\ell$ is small -- say $\ell=o(n)$ -- then the distribution $D_{\sigma}$ is far from uniform in $\{0,1\}^n$, as it is supported on a set of size $\leq 2^\ell$. 
Nonetheless, we will now show that $D_\sigma$ is in fact pseudorandom -- i.e., it is indistinguishable from uniform to an adversary, even if that adversary is equipped with black box access to $\sigma$ and $\sigma^{-1}$.

Before we prove this, we claim that this is a first step towards proving our quantum pseudorandomness conjecture. In particular, this can be seen as a classical model of our quantum pseudorandomness construction in which the random $\sigma$ is analogous to the evolution $U$ of the CFT by the scrambling time (or a multiple thereof), and random choice of $X$ is analogous to a random Pauli insertion. 

The main theorem of the section is as follows:

\begin{theorem}
\label{thm:classicalPRproperty}
Pick a random $\sigma$ from $S_{2^n}$ and let $\ell=o(n)$.
Suppose that one is given a string $y$ which is either drawn from $D_{\sigma}$, or else was drawn uniformly\footnote{We note that the support of $D_\sigma$ is of size at most $2^\ell$, so these two possibilities are nearly perfectly distinct from one another.} from $\{0,1\}^n$. 
Then, given $y$ and black-box access to $\sigma$ and $\sigma^{-1}$, no quantum algorithm can determine which is the case with probability $>2/3$ with fewer than $2^{\Omega(\ell)}$ quantum queries to $\sigma$ and $\sigma^{-1}$.
\end{theorem}
Here when we say ``with probability $\geq 2/3$'', this probability is taken over the choice of $\sigma$, the choice of $y$ from $D_{\sigma}$, and the randomness of the quantum algorithm -- so we are claiming an \emph{average-case} lower bound for this problem.
Also we are assuming that $\sigma$ and $\sigma^{-1}$ are queried as standard classical oracles, i.e. one has access to $\mathcal{O}_{\sigma} : \ket{x}\ket{y} \rightarrow \ket{x}\ket{y\oplus \sigma(x)}$, and that one can query this map $\mathcal{O}_\sigma$ (or $\mathcal{O}_{\sigma^{-1}}$)in superposition.
This immediately implies the same result if one only has access to the ``in place'' oracle $\mathcal{IP}_{\sigma} : \ket{x} \rightarrow \ket{\sigma(x)}$ (or analogously $\mathcal{IP}_{\sigma^{-1}}$), since one can simulate queries to the in-place oracle by querying $\mathcal{O}_\sigma$ and $\mathcal{O}_{\sigma^{-1}}$ as standard classical oracles\footnote{To simulate $\mathcal{IP}_\sigma$ on $\Sigma_x \alpha_x \ket{x}$, one simply applies $\mathcal{O}_\sigma$ to obtain $\Sigma_x \alpha_x \ket{x}\ket{\sigma(x)}$, and then applies $\mathcal{O}_{\sigma^{-1}}$ from the second to the first register to erase the ``garbage'' qubits to obtain $\Sigma_x \alpha_x \ket{\sigma(x)}$ in tensor product with $\ket{0^n}$}.

\begin{proof}

Our theorem will be proven via a reduction from an average-case variant of unstructured search. From there, a query lower bound can readily be established by invoking a standard hybrid argument \cite{BBBV}.

Let us now describe this reduction. For simplicity, we will first consider the case that the algorithm can only query $\sigma$ but not $\sigma^{-1}$. We later generalize to the case one can query both $\sigma$ but not $\sigma^{-1}$.
We can think of $\sigma$ as defining a tree $T(\sigma)$ where each node is labeled by a string in $\{0,1\}^n$. The root of the tree is $0^n$, the left child of node $x$ is $\sigma(x)$ and the right child is $\pi(x)$. The depth of the tree is $\ell$.
Let $L(\sigma)$ denote the leaves of the tree and let $R(\sigma)$ denote
the subset of strings in the second to last row of the tree -- i.e. the nodes just above $L(\sigma)$.
See Figure \ref{fig:sigmatree} for a pictoral representation of these sets.
Note that since the depth $\ell=o(n)$, with high probability over the choice of $\sigma$, all strings in the tree will be distinct -- so with high probability $|L(\sigma)|=2^\ell$ and $|R(\sigma)|=2^{\ell-1}$.

Sampling from the distribution $D_{\sigma}$ is equivalent to outputting a uniformly random leaf $y$ of the tree $T(\sigma)$.
Therefore, $y$ is a uniformly random element of $L(\sigma)$ and a uniformly random child of $R(\sigma)$.
In the other case -- when $y$ is independently drawn form the uniform distribution -- $y$ is not in $L(\sigma)$ with very high probability. 
Therefore, we claim that the problem of distinguishing these distributions is as hard as the following search problem: 
given black-box access to $\sigma$ as well as a list of strings $R(\sigma)$, does there exist an $x\in R(\sigma)$ such that either $\sigma(x)=y$ or $\pi(x)=y$?

This is clearly a variant of unstructured search, and we claim that a $2^{\Omega(\ell)}$ queries to $\sigma$ are required to solve this problem, even in the average case.
This follows because the search lower bound of \cite{BBBV} is not only a bound for worst-case search, but also a bound for \emph{average-case} search, in which the marked item is placed at a uniformly random location.
Analogously one can obtain an average-case lower bound for our problem of interest from a hybrid argument -- that is for the case when $y$ is either a random child of a uniformly random element of $R(\sigma)$ or else does not appear as a child of any element of $R(\sigma)$.

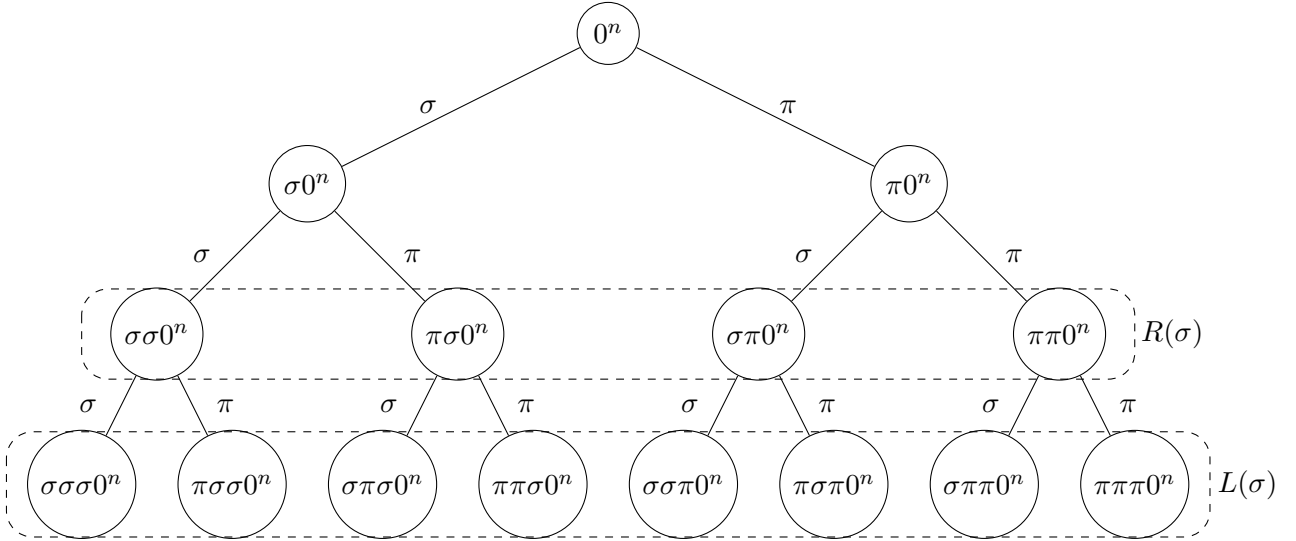
\begin{figure}
    \centering
    \begin{tikzpicture}[level distance=2cm,
  level 1/.style={sibling distance=8cm},
  level 2/.style={sibling distance=4cm},
  level 3/.style={sibling distance=2cm},
  nodes={draw, circle}, minimum size=1.5em]
  \node {$0^n$}
    child {node {$\sigma 0^n$}
      child {node(1) {$\sigma \sigma 0^n$}
        child{ node(3) {$\sigma\sigma\sigma0^n$}
        edge from parent node[left,xshift=-3,draw=none]{$\sigma$}}
        child{ node {$\pi\sigma\sigma0^n$}
        edge from parent node[right,xshift=3,draw=none]{$\pi$}}
      edge from parent node[left,xshift=-3,draw=none]{$\sigma$}}
      child {node {$\pi \sigma 0^n$} 
        child{ node {$\sigma\pi\sigma0^n$}
        edge from parent node[left,xshift=-3,draw=none]{$\sigma$}}
        child{ node {$\pi\pi\sigma0^n$}
        edge from parent node[right,xshift=3,draw=none]{$\pi$}}
      edge from parent node[right,xshift=3,draw=none]{$\pi$}}
      edge from parent node[left,xshift=-3,draw=none]{$\sigma$}
    }
    child {node {$\pi 0^n$}
    child {node {$\sigma \pi 0^n$} 
        child{ node {$\sigma\sigma\pi0^n$}
        edge from parent node[left,xshift=-3,draw=none]{$\sigma$}}
        child{ node {$\pi\sigma\pi0^n$}
        edge from parent node[right,xshift=3,draw=none]{$\pi$}}
    edge from parent node[left,xshift=-3,draw=none]{$\sigma$}}
      child {node(2) {$\pi \pi 0^n$} 
        child{ node {$\sigma\pi\pi0^n$}
        edge from parent node[left,xshift=-3,draw=none]{$\sigma$}}
        child{ node(4) {$\pi\pi\pi0^n$}
        edge from parent node[right,xshift=3,draw=none]{$\pi$}}
      edge from parent node[right,xshift=3,draw=none]{$\pi$}}
      edge from parent node[right,xshift=3,draw=none]{$\pi$}
    };
    \draw[dashed,rounded corners=10]($(1) + (-1,0.6)$)rectangle($(2) +(1,-0.6)$);
    \draw[dashed,rounded corners=10]($(3) + (-1,0.7)$)rectangle($(4) +(1,-0.7)$);
    \node[draw=none] at (7.5,-4) {$R(\sigma)$};
    \node[draw=none] at (8.5,-6) {$L(\sigma)$};
\end{tikzpicture}

    \caption{Representation of the tree $T(\sigma)$ for $k=3$. $R(\sigma)$ denotes the nodes in the second to last row of the tree, and $L(\sigma)$ denotes the leaves of the tree. }
    \label{fig:sigmatree}
\end{figure}

However, there is an important subtlety to address in order to apply these lower bounds to the problem at hand. 
Namely, if $y$ were completely uncorrelated with $\sigma$, then this \emph{would} be straightforward unstructured search, and the BBBV lower bound would immediately hold. 
However, in our case $y$, $\sigma$ and $R(\sigma)$ are not totally independent of one another in the ``yes'' case. In particular there is a slight correlation between $y$, $\sigma$ and the elements of $R(\sigma)$.
Furthermore $\sigma$ has some additional structure -- it is restricted to be a permutation -- which prevents it from being set to an arbitrary function. 
So this reduction does not straightforwardly go through.

Therefore to complete our proof, we need to quantify this correlation, and show it is insignificant.
The basic reason this is true is that, for this low of a value of $\ell$ (i.e. for this depth of the tree),
we have fixed very few entries of the permutation. 
Therefore the sets $R(\sigma)$ and $y$ are close to being independent of one another.
Furthermore at this low of depth of the tree, $\sigma$ is indistinguishable from a random function, as has been shown by Yuen \cite{yuenrandomfunction}. 
At a high level these observations are analogous to \cite{BBBV}'s lower bound for permutation inversion, which also required showing that some additional structure in the problem is irrelevant to the lower bound.

To show this more rigorously for our setup, we will need to define a few hybrid distributions.
Let $S=\{\sigma \in S_{2^n} : T(\sigma) \text{ has all distinct elements}\}$. It is easy to see that $S$ is a $1-1/\text{exp(n)}$ fraction of $S_{2^n}$, because the depth of the trees considered here are so low. Consider the following distributions on inputs
\begin{itemize}
    \item (A) Draw $\sigma$ uniformly from $S_{2^n}$ and $y$ uniformly and independently from $\{0,1\}^n$
    
    \item (B) Draw $\sigma$ uniformly from $S$ and $y$ uniformly and independently from $\{0,1\}^n - T(\sigma)$ -- i.e. $y$ is drawn from the complement of all strings appearing in the tree $T(\sigma)$.
    \item (C) Draw $\sigma'$ uniformly from $S$ and $y$ uniformly and independently from $\{0,1\}^n-T(\sigma')$. Now pick $x$ uniformly from $L(\sigma')$ and set $\sigma= SWAP(x,y) \circ \sigma'$. In other words, pick both $\sigma'$ and $y$ independently, and then let $\sigma$ be the slight modification of $\sigma'$ to have $y$ in the last row of its tree. Here $SWAP(x,y)$ is the permutation that exchanges strings $x$ and $y$ while leaving all other strings fixed. Note that by picking $y$ from outside the tree, we have ensured that $\sigma \in S$. 
    
    \item (D) $\sigma$ is drawn uniformly from $S$ and $y$ is drawn uniformly from $L(\sigma)$.
    
    \item (E) $\sigma$ is drawn uniformly from $S_{2^n}$ and $y$ is drawn uniformly from $L(\sigma)$.
    
\end{itemize}

Our goal is to show that distributions (A) and (E) are indistinguishable from one another. By the standard hybrid argument it suffices to show (A) is indistinguishable from (B), (B) from (C), etc.

First note that that distribution (A) is inverse exponentially close in total variation distance to distribution (B) -- this follows from the fact that $S$ is of large size in $S_{2^n}$, and that the number of strings outside $T(\sigma)$ namely ($2^n-O(2^\ell)$) is also of large size in $\{0,1\}^n$.
Similarly distributions (D) and (E) are inverse exponentially close in total variation distance as well. Therefore we simply need to show that (B) is indistinguishable from (C), and (C) is indistinguishable from (D). 

To see why (B) is indistinguishable from (C) a standard hybrid argument suffices. The only difference between these distributions over strings is whether only two entries of $\sigma$ have been changed -- namely that $\sigma(\sigma^{-1}(x))$, which used to be some value $x$ at a uniformly random leaf of $T(\sigma),$ has been replaced with $y$ and vice versa. 
Therefore one can use a standard hybrid argument\footnote{Consider the average query amplitude on elements of $R(\sigma)=R(\sigma')$. Since $\sigma\in S$ we have that $|R(\sigma)|=2^{\ell-1}$, so the average query magnitude on any single entry is at most $2^{-(\ell-1)/2}$. 
One then applies a hybrid argument, where the hybrids are replacing $\sigma$ with $\sigma'$ at each step in the algorithm.} to lower bound the number of queries to $\sigma$ needed to determine if this switch has been made. The lower bound obtained is that $\Omega(2^{\ell/2})$ queries to $\sigma$ are required to distinguish these distributions. 
This lower bound even holds against this average-case problem, because it is a form of search where the marked item is at a uniformly random location.

Therefore all that remains to be done is to show that distributions (C) and (D) are indistinguishable. This follows because, in fact, these are the same distribution! To see this, consider the probability of seeing a certain pair $y,\sigma$ in each distribution. Clearly in distribution (D) we have that
\begin{equation}
    \Pr_D[y,\sigma] =  
    \begin{cases}
      \frac{1}{|S| 2^\ell}, & \text{if}\ y\in L(\sigma) \text{ and } \sigma\in S \\
      0, & \text{otherwise}
    \end{cases}
  \end{equation}

Now let us compute $\Pr_C[y,\sigma]$. Since by construction $\sigma,y$ are drawn such that $\sigma\in S$ and $y\in L(\sigma)$, this probability will be zero if these conditions are not met, as in probability distribution $D$.
So we only need to compute $\Pr_C[y,\sigma]$ in the case that $\sigma\in S$ and $y\in L(\sigma)$. In particular, it is easy to see that for any $\sigma\in S$ and any $y\in L(\sigma)$, and any $\tau\in S$ and any $w\in L(\tau)$, that
\[\Pr_C[\sigma,y]=\Pr_C[\tau,w].\]

This follows by direct calculation. To see this, one can explicitly compute this probability by the chain rule (assuming that $\sigma\in S$ and $y\in L(\sigma)$):
\begin{align*}
\Pr_C[\sigma,y] &= \displaystyle\sum_{x \in \{0,1\}^n,\sigma' \in S} \Pr[\sigma'] \Pr[y|\sigma'] \Pr[x|\sigma',y] \Pr[\sigma|\sigma,y,x] \\
&= \frac{1}{|S|}\frac{1}{2^n - |T(\sigma)|} \frac{1}{2^\ell} \displaystyle\sum_{x \in \{0,1\}^n,\sigma' \in S}  \Pr[\sigma|\sigma,y,x] \\
&=\frac{\text{\# of pairs }\sigma',x \text{ such that }\sigma=SWAP(x,y) \circ \sigma' }{|S|2^\ell(2^n-|T(\sigma)|)} \\
&= \frac{1}{|S|2^\ell}
\end{align*}
Where the first line follow from the chain rule, the second and third follow from direct calculation, and the last follows from the fact that there is one such pair for each $x\in \{0,1\}^n \ T(\sigma)$.
Hence hybrids (C) and (D) are identical. Therefore by the standard hybrid argument we can conclude that $2^{\Omega(\ell/2)}$ queries to $\sigma$ are required to distinguish distributions (A) and (E).

This completes the proof in the case where one can query $\sigma$ only, but not $\sigma^{-1}$.
An analogous proof holds when one can query $\sigma^{-1}$ but not $\sigma$. To see this, consider the tree obtained by applying $\sigma^{-1}$ or $\pi^{-1}$ at each step starting from $y$.
In other words, consider building the tree \emph{upwards} from the leaf $y$, applying $\sigma^{-1}$ or $\pi^{-1}$ at each step.
Distinguishing these distributions is equivalent to determining whether or not $0^n$ is a leaf of this upward-directed tree.
By the same proof, one can prove that one needs $\Omega(2^{\ell/2})$ queries to $\sigma^{-1}$ to distinguish these distributions.

To generalize to the case where one can query both $\sigma$ and $\sigma^{-1}$, one can simply combine the above proofs. Consider choosing $\sigma$ and $y$ independently randomly -- with high probability $y$ will not appear in the tree $T(\sigma)$. Now consider building the tree obtained by applying $\sigma,\pi$ to root $0^n$ to height $h=\lfloor \ell/2 \rfloor -1$, and build the ``reversed'' tree upwards from the leaf $y$ by applying $\sigma^{-1},\pi^{-1}$ to height $\ell-h-1$ (see Figure \ref{fig:sigmaweldedtree}). Call the sets of leaves of these trees $A,B$ respectively. 
Determining if $y\in L(\sigma)$ is equivalent to determining if $\exists a\in A \exists b\in B: \sigma(a)=b$ -- or equivalently if $\exists a\in A \exists b\in B: \sigma^{-1}(b)=a$.
Critically now there is a certain symmetry between $\sigma$ and $\sigma^{-1}$; the problem reduces to search over either oracle.
Now, analogously to the previous proof, consider the hybrid distribution obtained by picking a random element of $a\in A$, a random element $b\in B$, and altering $\sigma$ by a single SWAP\footnote{Specifically, set $\sigma \rightarrow  SWAP(\sigma(a),b)\circ \sigma$} such that $\sigma(a)=b$. 
Under this distribution, $y\in L(\sigma)$ by construction -- and similarly one can show it is close in total variation distance to the distribution $D_{\sigma}$, again by symmetry properties of the distributions.
But these distributions only differ by a random single entry of $\sigma$ (correspondingly, $\sigma^{-1}$), so again a hybrid argument gives a lower bound of $\Omega(2^{\ell/2})$ queries to $\sigma$ or $\sigma^{-1}$ for distinguishing these distributions (as there are $\Omega(2^{\ell})$ different pairs $a\in A, b\in B$ one could choose to swap in the above argument). This completes the proof of the lower bound.

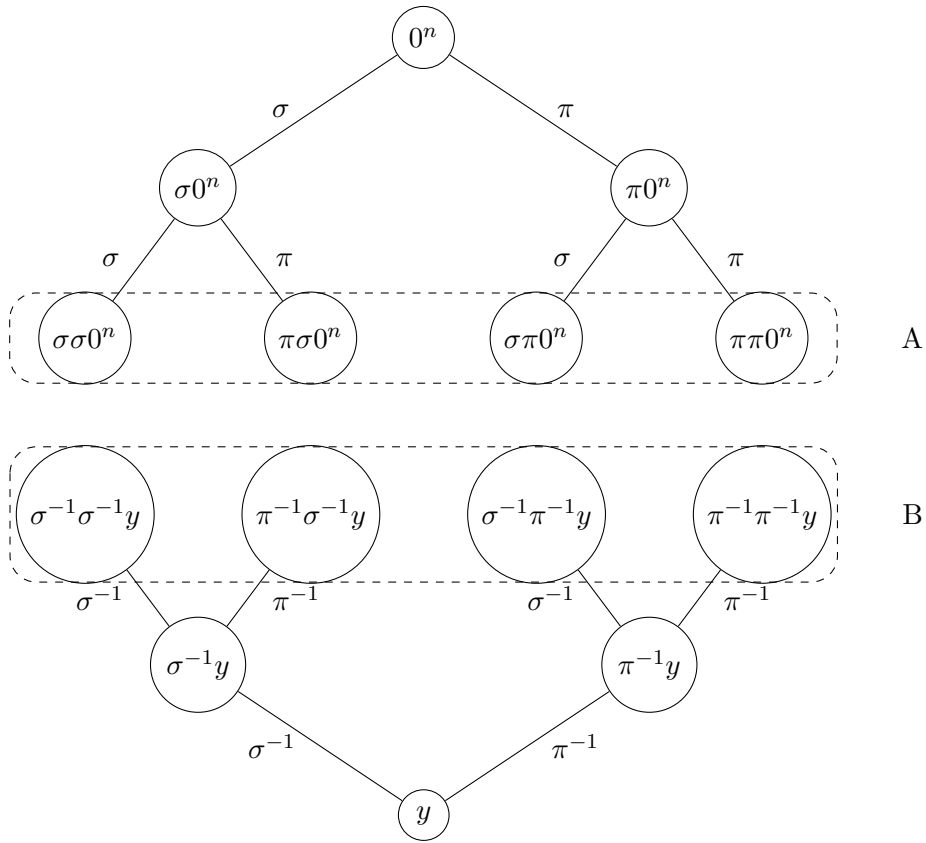
\begin{figure}
    \centering
    \begin{tikzpicture}[level distance=2cm,
  level 1/.style={sibling distance=6cm},
  level 2/.style={sibling distance=3cm},
  level 3/.style={sibling distance=1.5cm},
  nodes={draw, circle}, minimum size=1.5em, grow=down]
  \node {$0^n$}
    child {node {$\sigma 0^n$}
      child {node(1) {$\sigma \sigma 0^n$}
      edge from parent node[left,xshift=-3,draw=none]{$\sigma$}}
      child {node {$\pi \sigma 0^n$} 
      edge from parent node[right,xshift=3,draw=none]{$\pi$}}
      edge from parent node[left,xshift=-3,draw=none]{$\sigma$}
    }
    child {node {$\pi 0^n$}
    child {node {$\sigma \pi 0^n$} 
    edge from parent node[left,xshift=-3,draw=none]{$\sigma$}}
      child {node(2) {$\pi \pi 0^n$} 
      edge from parent node[right,xshift=3,draw=none]{$\pi$}}
      edge from parent node[right,xshift=3,draw=none]{$\pi$}
    };
     \draw[dashed,rounded corners=10]($(1) + (-1,0.6)$)rectangle($(2) +(1,-0.6)$);
     \node[draw=none] at (6.5,-4) {A};
\end{tikzpicture}

   \vspace{2em}

    \begin{tikzpicture}[level distance=2cm,
  level 1/.style={sibling distance=6cm},
  level 2/.style={sibling distance=3cm},
  level 3/.style={sibling distance=1.5cm},
  nodes={draw, circle}, minimum size=1.5em, grow'=up]
  \node {$y$}
    child {node {$\sigma^{-1} y$}
      child {node(1) {$\sigma^{-1} \sigma^{-1} y$}
      edge from parent node[left,xshift=-3,draw=none]{$\sigma^{-1}$}}
      child {node {$\pi^{-1} \sigma^{-1} y$} 
      edge from parent node[right,xshift=3,draw=none]{$\pi^{-1}$}}
      edge from parent node[left,xshift=-3,draw=none]{$\sigma^{-1}$}
    }
    child {node {$\pi^{-1} y$}
    child {node {$\sigma^{-1} \pi^{-1} y$} 
    edge from parent node[left,xshift=-3,draw=none]{$\sigma^{-1}$}}
      child {node(2) {$\pi^{-1} \pi^{-1} y$} 
      edge from parent node[right,xshift=3,draw=none]{$\pi^{-1}$}}
      edge from parent node[right,xshift=3,draw=none]{$\pi^{-1}$}
    };
     \draw[dashed,rounded corners=10]($(1) + (-1,0.9)$)rectangle($(2) +(1,-0.9)$);
     \node[draw=none] at (6.5,4) {B};
\end{tikzpicture}

    \caption{Generalization of the proof to the case one can query both $\sigma$ and $\sigma^{-1}$ for $\ell=5$. By choosing whether or not to connect these trees at a random pair of points $a\in A,b\in B$, one changes a ``yes'' instance to a ``no'' instance. A hybrid argument then proves the lower bound on distinguishing these two cases.}
    \label{fig:sigmaweldedtree}
\end{figure}

\end{proof}

\subsection{Sketch of a quantum black-box pseudorandomness proof}

\label{sec:quantumpseudorandomness}

We will now explain why a straightforward generalization of these results applies to a quantum variant of the model. To recap, imagine that we have black-box access to a Haar-random unitary $U$ as well as its inverse. and at $\ell$ different times we insert a randomly chosen Pauli between applications of $U$. Let $k\in\{I,X,Y,Z\}^\ell$ denote the secret key and let
\[\ket{\Psi_k} = k_\ell U k_{\ell-1} U \ldots U k_1 U \ket{0^n}\]
We now want to show this forms a PRS ensemble for say $\ell=n^\delta$ (for some small $\delta>0$) if one only has black box-access to $U$ and $U^{-1}$ (Claim \ref{theorem:quantumpseudorandomness}).

To do so, we can follow the proof of Theorem \ref{thm:classicalPRproperty}.
Consider the analogous tree of states $T(U)$ of depth $k$ -- where the root is labelled by $\ket{0^n}$, and each node $\ket{\phi}$ has 4 children containing states $U\ket{\phi}$, $X_1 U\ket{\phi}$, $Y_1 U\ket{\phi}$, and $Z_1 U\ket{\phi}$.
Once again let $L(U)$ denote the states at the leaves of the tree and $R(U)$ denote the states in the second to last row of the tree. 

We now sketch the proof of this quantum generalization of Theorem \ref{thm:classicalPRproperty} in the case one can only query $U$; the generalization to querying both $U$ and $U^{-1}$ follows analogously. 

First we claim that with high probability over the choice of $U$, all states in the tree $T(U)$ are nearly orthogonal to one another (let ``nearly orthogonal'' mean the norm of their inner product is $\leq 2^{-O(n)}$) with high probability.
This follows from a tedious but straightforward application of the well-known formulae for integrals over the Haar measure \cite{collins2006integration}.
This is easy to shown for $\ell=1$ -- for instance one can easily show that
\[\mathbb{E}_{U} \left|\bra{0^n} U^\dagger X_1 U \ket{0^n}\right|^2 =\frac{2^{n}}{2^{2n}-1} - \frac{1}{2^{2n}-1} = O\left(\frac{1}{2^n}\right) \]
using explicit formulae for integration over the Haar measure (see e.g. \cite{BBJsupergrover} SI Section D).
In other words, applying a Pauli to a Haar random state $\psi$ renders it nearly orthogonal to itself.
Therefore for $\ell=1$ the states $\ket{\psi_k}$ for $k=I$ or $k=X$ are nearly orthogonal with probability $\geq 1-2^{-O(n)}$ by Markov's inequality.
Expanding this to the general case of higher $\ell$ is considerably more tedious. 
For instance, suppose that one wishes to show that $XU^\ell \ket{0^n}$ and $U^\ell \ket{0^n}$ are nearly orthogonal with high probability (again for $\ell=O(n^\delta)$). 
This can be obtained using the explicit formula for $\ell$-th moments of the Haar measure from Collins and Sniady \cite{collins2006integration}:
\begin{align*}\mathbb{E}_U U_{i_1j_1}U_{i_2 j_2} \ldots U_{i_\ell j_\ell} U^{\dagger}_{i'_1i'_2}\ldots U^{\dagger}_{i'_\ell j'_\ell} =   \displaystyle\sum_{\sigma,\tau \in S_\ell} \delta_{i_1 i'_{\sigma(1)}} \ldots \delta_{i_\ell i'_{\sigma(\ell)}} \delta_{j_1 j'_{\tau(1)} } \ldots \delta_{j_\ell j'_{\tau(\ell)}} Wg(\sigma \tau^{-1}, 2^n)\end{align*}
where $Wg(\sigma, d)$ denotes the Weingarten function in dimension $d$.
Using this formula one can show that indeed the expected norm of the inner product between $U^\ell \ket{0^n}$ and $X_1 U^\ell\ket{0^n}$ is exponentially small in $n$ -- we give a proof of this in Appendix \ref{app:orthogonality}.
We expect the same result would apply for $\ket{\Psi_k}$ for any two distinct secret keys $k$ by a tedious application of the Haar integration functions, but we do not prove this fact.
Once this is established, the claim that all elements of the tree are nearly orthogonal with high probability follows from a union bound.

Now that the orthogonality property has been established for typical $U$, we now define hybrid distributions analogous to those defined in the proof of Theorem \ref{thm:classicalPRproperty}.
Let $S$ be the set of unitaries for which all nodes of the tree $T(U)$ are approximately orthogonal. By the above argument $S$ has large measure under the Haar measure. 
By analogous arguments to the classical case the argument boils down to showing the following three distributions are indistinguishable:

\begin{itemize}

    \item (B) Draw $U$ uniformly from $S$ and $\ket{\psi}$ uniformly and independently from $\mathbb{C}^{2^n}$. With high probability $\ket{\psi}$ will be nearly orthogonal to all states in $T(U)$.

    \item (C) Draw $U'$ uniformly from $S$ and $\ket{\psi}$ uniformly and independently from $\mathbb{C}^{2^n}$.  
    Now pick $\ket{\phi}$ uniformly from $L(U')$ and set $U= SWAP(\ket{\phi},\ket{\psi}) \circ U'$. In other words, pick both $U'$ and $\ket{\psi}$ independently, and then let $U$ be the slight modification of $U'$ to have $\ket{\psi}$ in the last row of its tree. Here $SWAP(\ket{\phi},\ket{\psi})$ is the unitary that exchanges the states $\ket{\phi}$ and $\ket{\psi}$ while acting as the identity on the remaining space. Note that since $\ket{\psi}$ is nearly orthogonal to all states in the tree almost surely, we have ensured that $U \in S$ with high probability. 
    
    \item (D) $U$ is drawn uniformly from $S$ and $\ket{\psi}$ is drawn uniformly from $L(U)$.
    
\end{itemize}

Once again one can show that these three distributions are indistinguishable with fewer than $2^{\Omega(\ell)}$ queries to $U$. In particular (B) and (C) are again indistinguishable by a standard hybrid argument -- indeed, if one knows the basis for $R(U)$, then this problem reduces to determining if we have appended $SWAP$ to the unitary starting from one of the $2^{\ell-1}$ elements of $R(U)$ -- i.e. we are asking if we have changed a single entry of the unitary $U'$ in this basis. This therefore reduces to a form of search, and the standard hybrid argument applies.

We also claim distributions (C) and (D) are close in total variation distance. Again this follows by a straightforward calculation: under the Haar measure on the product space of $U(d) \times \mathbb{C}^{2^n}$, the PDF on a pair $(U,\ket{\psi})$ under distribution (D) is
\[\Pr_D[U,\ket{\psi}] = \begin{cases}
      \frac{c}{2^\ell} \delta(\ket{\psi}), & \text{if}\ U\in S \text{ and } \ket{\psi} \text{ is the first leaf in } L(U) \\
      \frac{c}{2^\ell} \delta(\ket{\psi}), & \text{if}\ U\in S \text{ and } \ket{\psi} \text{ is the second leaf in } L(U) \\
      \vdots \\
      \frac{c}{2^\ell} \delta(\ket{\psi}), & \text{if}\ U\in S \text{ and } \ket{\psi} \text{ is the last leaf in } L(U) \\
      0, & \text{otherwise}
    \end{cases} \]
where $c$ is the constant such that the indicator function of $U\in S$ integrates to $1/c$ over the Haar measure on $U(d)$ (so $c$ is $1-1/\text{exp(n)}$), and $\delta$ is the Dirac delta function on $\mathbb{C}^{2^n}$.

Similarly one can compute the probability of these events under distribution (C); clearly by construction the $\ket{\psi}$ is always a leaf in $L(U)$ (with a uniform distribution on its position within $L(U)$, so this probability distribution must have the same support. 
Therefore one must simply show that all of these options are equally likely. This again should follow from a straightforward calculation using the chain rule.

This completes the proof in the case one can query $U$ only, and a straightforward generalization to querying both $U,U^{-1}$ should follow analogously to the classical case.

Finally, we note that there is one additional step required for the quantum lower bound which is absent from the classical case.
Namely, since quantum states cannot be copied, one needs to consider the problem of deciding whether $\ket{\psi}$ was drawn from $D_U$, or else Haar random, \emph{given multiple copies} of $\ket{\psi}$.
Fortunately our lower bound is robust against this modification -- as our query lower bound is coming from the classical lower bound for search, it holds even if one knows an exact description of $\ket{\psi}$ (and of the basis of $R(U)$). Therefore the lower bound of $2^{\Omega(\ell)}$ queries holds in this model as well.

\subsection{Further physical considerations for our pseudorandomness model}

\label{sec:energyissues}
We have shown that, if the only way one ``accesses'' the Hamiltonian is by black-box time evolution with the unitary $U=e^{-i\hamilcft m t_{scr}}$, then one cannot distinguish the states in Claim \ref{theorem:quantumpseudorandomness} from Haar-random states with few queries to $U$.

However, in the case of the AdS/CFT setup, one could consider different forms of access to the dynamics of the system which might allow one to break this statement.
For instance, suppose that one had a complete description of the local Hamiltonian $\hamilcft$ of the CFT. 
This would allow one to measure the energy of a state $\ket{\psi}$ under this Hamiltonian, i.e. $\left|\bra{\psi}\hamilcft\ket{\psi}\right|^2$, by measuring each term of the Hamiltonian in sequence, with polynomially many copies of the state.
This energy measurement would falsify the above statement; for instance the TFD state has relatively low energy by construction compared to a Haar-random state, since time evolution by $\hamilcft$ does not change the energy of the system, and since each perturbation by a local Pauli only adds an $O(1)$ amount to the energy of the system, the states in $F_{K,T}$ used in our pseudorandomness construction will have low energy. 
Indeed, this is precisely why we chose such states -- because these low energy CFT states are precisely the states with smooth gravity duals, for which the complexity duality conjectures makes sense. 
In contrast a Haar-random state will have relatively high energy with high probability. 
Therefore, allowing energy measurements on an arbitrary polynomial number of copies of the state enables a distinguisher algorithm to break our construction.

While this is true, we claim this need not affect the conclusions of our paper.
First, proving the pseudorandomness of our construction is not actually necessary for our argument to hold in the first place. The only aspect of pseudorandomness that we use is that different versions of this construction -- with different lengths of time evolution corresponding to different wormhole volumes-- are indistinguishable from one another.
The use of a Haar-random state in the argument is only an intermediate hybrid. 
Therefore, we only need to argue that different states in $F_{K,T}$ -- which all have low energy -- are directly indistinguishable from one another.

Second, as previously noted in Conjecture \ref{conj:pseudorandomness}, we do not need to prove indistinguishability against arbitrary polynomial numbers of copies of the state -- rather only against the number of copies $c(n)$ needed to very roughly approximate the volume of the wormhole in the AdS side. If this is achievable with say a constant number of states, then this limited number of states would make it impossible to perform even a single energy measurement to any degree of accuracy. 
Even if say $O(n)$ states are needed to distinguish these wormhole volumes on the AdS side, the precision of the energy measurement required to distinguish these states in the CFT would be a prohibitively large polynomial.  

We claim that certain variants of our construction can foil this distinguisher.
To see this concretely, note that there are many variants of our PRS construction to instantiate wormholes of different sizes in our arguments.
For instance, one could fix $m$ -- the number of multiples of the scrambling time between operator insertions -- and vary $\ell$ -- the number of insertions -- to vary the length of the wormhole, which is $O(m\ell t_{scr})$. 
This might still ``give away'' the wormhole length with few energy measurements though -- because the number of operator insertions, and hence the energy, would vary between the long and short wormholes, and hence a single energy measurement might break the construction.
Alternatively one could fix $\ell$ and vary $m$ to vary the length of the wormhole
Again this would ``give away'' the wormhole length with few energy measurements, because the spacing between the last two insertions varies between the long and short wormholes, and is relatively easy to measure with few energy measurements.
To fix this issue, note that one also has the freedom to apply the shocks to different qubits, rather than just the first qubit of the ensemble, as we did in our construction for simplicity of explanation. Furthermore, one can apply them at randomly chosen times, rather than at fixed spacings.
Therefore, suppose that we insert the random shocks on randomly chosen qubits at each step, and that we apply them at randomly chosen times, with varying lengths $T$ of total time evolution to vary the wormhole length. How many copies of the state our be needed to break our construction with energy measurements? Intuitively this would force the adversary to perform \emph{local} energy density measurements at each of the $n$ qubits, and at each time step, to a certain degree of accuracy to reveal when/where the operator insertion occurred, therefore allowing them to reverse-engineer the insertions and learn the length of time evolution of the system. This requires measuring the energy in $n$ locations at $T$ timesteps to a certain level of precision\footnote{In particular, one would expect to need to measure the energy to sufficient accuracy to pin down the operator inserted with probability at least $1-\frac{1}{3\ell}$, to ensure that a union bound ensures all of the $\ell$ insertions are correctly deduced with probability $\geq 2/3$ by a union bound.}, requiring strictly more than $O(nT)$ copies of the state. In contrast we will argue in the next section that $O(1)$ copies of the state, or at most $O(T)$ copies (depending on one's precise views on what information the AdS state encodes), suffice for measuring the volume.
Hence this particular energy measurement attack on our construction requires too many copies of the state to affect our conclusion.

In short, since our pseudorandomness property is stronger than is necessary for our application, we conjecture that variants of our construction can foil attempts to break the argument\footnote{Unfortunately proving such indistinguishable rigorously in this model where the entire Hamiltonian is known is beyond the current reach of theoretical computer science. This is because the description of the Hamiltonian also gives a complete description of the circuit generating these states -- so proving indistinguishability would be equivalent to proving the unconditional security of a form of cryptography (which is equivalent to proving circuit lower bounds or separating $\mathsf{P}$ from $\mathsf{NP}$). }, since there should be some way to ``scramble'' the CFT state in the spirit of block ciphers to foil energy measurement attacks.
We leave pinning down this argument more precisely for future work (see Section \ref{sec:openproblems}).

\section{Volume is ``feelable''}
\label{sec:volumeeasy}

Let us describe why we believe the wormhole volume is efficiently approximable.
As we only need to distinguish large ($O(n^3)$ volume) vs. small\footnote{We note that our arguments would also work with smaller sized wormholes -- e.g. $n^{1+\delta}$ vs $n^{1+2\delta}$ volume for some small $\delta>0$ -- by setting $\ell=n^{\delta}$ vs $n^{2\delta}$ and $m=O(1)$ in Section \ref{sec:customPRS}.} ($O(n^2)$ volume) wormholes, even a coarse estimate suffices for our arguments.

There are several different notions of what it could mean for wormhole volume to be efficiently approximable.
The most basic of these would be to require that observers living in a copy of the AdS space could estimate the wormhole volume with local experiments. 
For instance, one could instantiate observers in the AdS space, designate one to be the ``referee'', and ask the referee output a guess if the wormhole is ``short'' or ''long'' after some polynomial amount of time.
If this guess is correct with decent probability, then one would say the wormhole volume is efficiently ``feelable.''

For our setting, it is unclear if wormhole volume is directly measurable in this way (see e.g. \cite{morris1988wormholes,bao2015wormhole}).
However, for our purposes, it suffices to argue that wormhole volume is feelable in a much more general setting.
This is because our pseudorandomness conjecture in the CFT is very strong -- it says that even if one had many copies of the CFT state, and could postprocess the results of experiments on these states with polynomial resources, then one could still not distinguish them from random. 
Therefore this means that even if wormhole volume is feelable in a more general setting -- where one can combine information about the experiences of many observers in the AdS space, possibly using several copies of the state, and then postprocess the results -- and/or if one can compute global properties of the AdS metric -- then this still has consequences for the AdS/CFT dictionary.

We argue that the wormhole volume is approximable in this more general model.
The basic idea is that the wormhole volume is a simple \emph{classical} property of the metric of the semi-classical AdS states under consideration.
If one's representation of the AdS system in the $n$-qubit AdS state allows one to efficiently extract this (classical) metric, then one could post-process the metric offline to compute the volume -- therefore indicating that even one copy of the AdS state would suffice.
Alternatively, if one views the given AdS state as only allowing access to one observer's experiences in the AdS at a time (say due to some strong form of complementarity \cite{susskind1993stretched}), then we claim that a wormhole of volume $T$ could be measured by simulating the experiences of several observers in the wormhole (and at most $O(T)$ would suffice). 
The basic idea is that a pair of observers in the AdS space can efficiently tell whether or not they are separated by a horizon -- by attempting to send a signal to one another, and seeing if the signal makes it.
That is, if one initializes two observers A and B at different points in the space-time, has A attempt to send a signal to B, time evolves this state by $H_{\text{AdS}}$, and measures the state at location B, one can tell if the signal was transmitted, and hence if A was separated from B by a horizon.
Therefore, the wormhole volume should be efficiently estimable by combining information about signals sent/received by many observers in the space, and postprocessing the results. 
There are several experiments one could consider to perform this task.
For instance, if one populates the space with many observers inside the wormhole, and collects information as to which pairs of observers were able to send signals to one another before crashing into the singularity, the number of successful signal transmissions would be a coarse estimate of the wormhole volume. 
Roughly speaking, the number of observers one would need to form a chain across the wormhole, in which each observer can communicate with the next before crashing into the singulaity, is roughly $O(T)$ -- because each observer has roughly order $\ell_{AdS}$ (the AdS curvature length, which we take to be constant) proper time before hitting the singularity, and the wormhole is of length roughly $T$.
So even if each copy of the state could only be used to simulate the experience of a single observer, $O(T)$ copies of the state would suffice.

In short, so long as one can time evolve the AdS state efficiently (that is, assuming the quantum Extended Church-Turing thesis), and/or one can efficiently extract the metric from the global AdS state, we can simulate this process in polynomial time on a quantum computer using relatively few ($O(1)$ or $O(T)$, depending on one's perspective) copies of the state. 
Therefore, given several copies of the AdS state, we argue that the volume is ``feelable'' to poly-time experiments.

\section{Discussion}
\label{sec:interpretation}

We have argued that any resolution of the wormhole growth paradox requires that either (A) the dictionary map is not computable in quantum polynomial time, (B) the quantum Extended Church-Turing thesis does not hold for quantum gravity.
There are many details to be pinned down to formalize these arguments -- see Section \ref{sec:openproblems}. Here we discuss what insights these conclusions might provide into the nature of the AdS/CFT correspondence. Our results are consistent with a number of different perspectives on the nature of AdS/CFT.

One possibility is that the AdS/CFT dictionary has high complexity. i.e. the mapping between the theories is exponentially complex\footnote{Note this does not necessarily imply the dual operators are themselves complex, only that it is very difficult to find them.}. This must hold even at early times in the CFT evolution, as our pseudorandomness arguments apply to quantum states which have evolved for only a polynomial amount of time. 
The dictionary or reconstruction map has been difficult to compute in practice, particularly behind the horizon (which is where the crux of our argument lies). 
Our results suggest that this could be due to the inherent complexity of the dictionary map, even at relatively early times in the evolution.
This is consistent with prior expectations that the dictionary may become increasingly complex to compute as one approaches the horizon \cite{susskind2014original}.
We note this conclusion holds either in a picture in which there exists a global AdS reconstruction map $\Phi$, or if there only exists one map $\Phi_P$ per causual patch $P$ (see Section \ref{sec:volumeeasy}) -- in which case our arguments would say that reconstructing some of the causual patches behind the horizon must be very complex.

We note a high complexity dictionary could limit the utility of the duality as a computational tool to understand properties of AdS states from properties of CFT states and vice versa.
Additionally, while AdS/CFT is a formal correspondence between a physical space and its dual space, certain tensor-network based interpretations of the duality may give one some form of access to the CFT state as well \cite{swingle2012entanglement,swingle2012constructing,pastawski2015holographic}.
The raises the possibility that, depending upon the nature of access to both the AdS and the CFT, a complex dictionary might mean that in this physical system one could perform computations more powerful than $\mathsf{BQP}$ -- because by perturbing the CFT state one could see the image of that perturbation in the AdS under a superpolynomially complex map.
Therefore a complex dictionary might produce a violation of the qECT as well. 
It would be interesting to further explore this line of thought.

Another possibility is to reject the quantum ECT in AdS, and conjecture that the dynamics of quantum gravity are exponentially complex to simulate even on a quantum computer.
Quantum computation is the first and only known theory that violates the classical ECT \cite{bv93}, and the quantum ECT formalizes the working assumption that this is the only violation -- an assumption which has been justified by results showing that even forms of quantum field theory are efficiently simulable on quantum computers \cite{jordan2012quantum,jordan2018bqp}.
If quantum gravity violates the qECT this would be remarkable, as it would be the first physical model with computational power beyond that of standard quantum computers.

Reconciling pseudorandomness with a low complexity dictionary and the qECT is more challenging -- it necessarily requires a fundamental modification of the AdS/CFT correspondence. 
The wormhole growth paradox demands a CFT quantity which is dual to volume which grows with time. 
Could there be a feelable quantity in the CFT which grows with time?
Unfortunately as sketched in Section \ref{sec:generalarg} the pseudorandomness arguments rule out the existence of such a quantity, because they not only show that complexity is not ``feelable'' in the CFT, but more generally that the length of time evolution from $\ket{TFD}$ to the CFT state is not ``feelable''. 
Thus, \emph{any} property of the quantum state alone which grows with time is not feelable in the CFT.

This suggests that the only way one can avoid the conclusion of a complex dictionary or dynamics is to modify the fundamental structure of AdS/CFT.
For instance, if one allows the AdS/CFT dictionary to depend on more than the state itself -- but say also some additional information about the CFT state -- then one can define analogues of complexity which are manifestly ``feelable'' and yet which are easy to compute.
We will describe such a quantity in Appendix \ref{sec:pseudocomplexity}, which we call ``pseudo-complexity.''

\section{Open problems}
\label{sec:openproblems}

In summary, the purpose of this paper was to provide evidence of pseudorandomness in the context of Complexity=Volume and spell out its consequences for AdS/CFT. 
While we presented evidence for pseudorandomness, there are several steps that must be taken to complete this formally, including rigorously showing our quantum lower bound in the case of a black-box Haar-random $U$, instantiating $U$ with a pseudorandom unitary construction, and formally pinning down the physical considerations of the model.
In addition there is work to be done to confirm and formalize the volume ``feelability'' argument.
We leave this open for future work.

Beyond the specific conclusions about the nature of AdS/CFT, the larger goal of this paper is to argue that computational considerations have much to illuminate about the nature of AdS/CFT and quantum gravity. As an example, for the purposes of our argument we did not need to describe the details of how to formalize the qECT for quantum gravity -- our arguments just use the qECT as a black box. However, it might be worth fleshing this out further, as well as pinning down more precisely in what sense the qECT is violated. This may provide further insights into the nature of AdS/CFT. 

In addition to the issue of resolving the central tension between computational pseudorandomness and the wormhole growth paradox, our work leaves open several additional problems:
\begin{itemize}
    
    \item One of the many reasons we could not use Ji, Liu and Song's PRS construction in our argument is that the quantum states produced do not obey the Ryu-Takayanagi formula \cite{ryu2006holographic}, nor do general random circuits such as those used in the work of \cite{brandao2016local}. A natural question is whether one can create a model of random circuits where the final state obeys the Ryu-Takayanagi formula with high probability. Such a circuit model could be taken as a toy model for the time evolution of the CFT state.
    
    \item What can our pseudorandomness construction say in light of Nielsen \emph{et al.}'s complexity geometry picture \cite{nielsen2006quantum}? Clearly one could formally state that finding short geodesics in this geometry is a very difficult computational problem, but is there more than one can say about the geometry itself?
    
    \item Does pseudo-complexity have a geometric interpretation along the lines of Nielsen's complexity geometry \cite{nielsen2006quantum}?
    
\end{itemize}

\section*{Acknowledgments}
We thank Scott Aaronson, Adam Brown, John Preskill, Douglas Stanford, Lenny Susskind, and Brian Swingle for detailed comments.
We also thank Chris Akers, Dorit Aharonov, Ahmed Almheiri, Ning Bao, Raphael Bousso, Matt DeCrosse, Helia Kamal, Dan Harlow, Patrick Hayden, Juan Maldacena, Saeed Mehraban, Dominik Neuenfeld, Sepher Nezami, Fabio Sanches, Jonah Sherman, Jalex Stark, Vincent Su, Michael Walter, and Ying Zhao for helpful discussions. 
A.B. and U.V. were supported in part by ARO Grant W911NF-12-1-0541, NSF Grant CCF-1410022, and a Vannevar Bush faculty fellowship. U.V. was supported by the Miller Institute at U.C. Berkeley through a Miller Professorship. 
B.F. acknowledges support from AFOSR YIP number FA9550-18-1-0148.



\begin{thebibliography}{ABDCV98}

\bibitem[Aar16]{aaronson2016complexity}
Scott Aaronson.
\newblock The complexity of quantum states and transformations: from quantum
  money to black holes.
\newblock {\em arXiv:1607.05256}, 2016.

\bibitem[ABDCV98]{aiello1998security}
William Aiello, Mihir Bellare, Giovanni Di~Crescenzo, and Ramarathnam
  Venkatesan.
\newblock Security amplification by composition: The case of doubly-iterated,
  ideal ciphers.
\newblock In {\em Annual International Cryptology Conference}, pages 390--407.
  Springer, 1998.

\bibitem[Ali15]{alishahiha2015holographic}
Mohsen Alishahiha.
\newblock Holographic complexity.
\newblock {\em Physical Review D}, 92(12):126009, 2015.

\bibitem[BB17]{bohdanowicz2017universal}
Thomas~C Bohdanowicz and Fernando~GSL Brand{\~a}o.
\newblock Universal {H}amiltonians for exponentially long simulation.
\newblock {\em arXiv:1710.02625}, 2017.

\bibitem[BBBV97]{BBBV}
Charles~H Bennett, Ethan Bernstein, Gilles Brassard, and Umesh Vazirani.
\newblock Strengths and weaknesses of quantum computing.
\newblock {\em SIAM journal on Computing}, 26(5):1510--1523, 1997.

\bibitem[BBJ16]{BBJsupergrover}
Ning Bao, Adam Bouland, and Stephen~P. Jordan.
\newblock Grover search and the no-signaling principle.
\newblock {\em Phys. Rev. Lett.}, 117:120501, 2016.

\bibitem[BHH16a]{brandao2016efficient}
Fernando~GSL Brand{\~a}o, Aram~W Harrow, and Micha{\l} Horodecki.
\newblock Efficient quantum pseudorandomness.
\newblock {\em Physical review letters}, 116(17):170502, 2016.

\bibitem[BHH16b]{brandao2016local}
Fernando~GSL Brandao, Aram~W Harrow, and Micha{\l} Horodecki.
\newblock Local random quantum circuits are approximate polynomial-designs.
\newblock {\em Communications in Mathematical Physics}, 346(2):397--434, 2016.

\bibitem[BM82]{blum1982generateconference}
Manuel Blum and Silvio Micali.
\newblock How to generate cryptographically strong sequences of pseudo random
  bits.
\newblock In {\em 23rd Annual Symposium on Foundations of Computer Science},
  pages 112--117. IEEE, 1982.

\bibitem[BPR15]{bao2015wormhole}
Ning Bao, Jason Pollack, and Grant~N Remmen.
\newblock Wormhole and entanglement (non-) detection in the {ER}={EPR}
  correspondence.
\newblock {\em Journal of High Energy Physics}, 2015(11):126, 2015.

\bibitem[BRS{\etalchar{+}}16a]{brown2016complexity}
Adam~R Brown, Daniel~A Roberts, Leonard Susskind, Brian Swingle, and Ying Zhao.
\newblock Complexity, action, and black holes.
\newblock {\em Physical Review D}, 93(8):086006, 2016.

\bibitem[BRS{\etalchar{+}}16b]{brown2016holographic}
Adam~R Brown, Daniel~A Roberts, Leonard Susskind, Brian Swingle, and Ying Zhao.
\newblock Holographic complexity equals bulk action?
\newblock {\em Physical review letters}, 116(19):191301, 2016.

\bibitem[BS19]{brakerski2019pseudo}
Zvika Brakerski and Omri Shmueli.
\newblock ({P}seudo) random quantum states with binary phase.
\newblock {\em arXiv:1906.10611}, 2019.

\bibitem[BV93]{bv93}
Ethan Bernstein and Umesh Vazirani.
\newblock Quantum complexity theory.
\newblock In {\em Proceedings of the Twenty-fifth Annual ACM Symposium on
  Theory of Computing}, STOC '93, pages 11--20, New York, NY, USA, 1993. ACM.

\bibitem[CHMP18]{chapman2018toward}
Shira Chapman, Michal~P Heller, Hugo Marrochio, and Fernando Pastawski.
\newblock Toward a definition of complexity for quantum field theory states.
\newblock {\em Physical review letters}, 120(12):121602, 2018.

\bibitem[CMR17]{carmi2017comments}
Dean Carmi, Robert~C Myers, and Pratik Rath.
\newblock Comments on holographic complexity.
\newblock {\em Journal of High Energy Physics}, 2017(3):118, 2017.

\bibitem[C{\'S}06]{collins2006integration}
Beno{\^\i}t Collins and Piotr {\'S}niady.
\newblock Integration with respect to the {H}aar measure on unitary, orthogonal
  and symplectic group.
\newblock {\em Communications in Mathematical Physics}, 264(3):773--795, 2006.

\bibitem[GGM86]{goldreich1986construct}
Oded Goldreich, Shafi Goldwasser, and Silvio Micali.
\newblock How to construct random functions.
\newblock {\em Journal of the ACM (JACM)}, 33(4):792--807, 1986.

\bibitem[HKL{\etalchar{+}}19]{harrow2019separation}
Aram~W Harrow, Linghang Kong, Zi-Wen Liu, Saeed Mehraban, and Peter~W Shor.
\newblock A separation of out-of-time-ordered correlator and entanglement.
\newblock {\em arXiv:1906.02219}, 2019.

\bibitem[HM18]{hackl2018circuit}
Lucas Hackl and Robert~C Myers.
\newblock Circuit complexity for free fermions.
\newblock {\em Journal of High Energy Physics}, 2018(7):139, 2018.

\bibitem[JKLP18]{jordan2018bqp}
Stephen~P Jordan, Hari Krovi, Keith~SM Lee, and John Preskill.
\newblock {BQP}-completeness of scattering in scalar quantum field theory.
\newblock {\em Quantum}, 2:Art--No, 2018.

\bibitem[JLP12]{jordan2012quantum}
Stephen~P Jordan, Keith~SM Lee, and John Preskill.
\newblock Quantum algorithms for quantum field theories.
\newblock {\em Science}, 336(6085):1130--1133, 2012.

\bibitem[JLS18]{ji2018pseudorandom}
Zhengfeng Ji, Yi-Kai Liu, and Fang Song.
\newblock Pseudorandom quantum states.
\newblock In {\em Annual International Cryptology Conference}, pages 126--152.
  Springer, 2018.

\bibitem[JM17]{jefferson2017circuit}
Robert~A Jefferson and Robert~C Myers.
\newblock Circuit complexity in quantum field theory.
\newblock {\em Journal of High Energy Physics}, 2017(10):107, 2017.

\bibitem[KC18]{kohler2018complete}
Tamara Kohler and Toby Cubitt.
\newblock Complete toy models of holographic duality.
\newblock {\em arXiv:1810.08992}, 2018.

\bibitem[LMPS16]{lehner2016gravitational}
Luis Lehner, Robert~C Myers, Eric Poisson, and Rafael~D Sorkin.
\newblock Gravitational action with null boundaries.
\newblock {\em Physical Review D}, 94(8):084046, 2016.

\bibitem[LR88]{luby1988construct}
Michael Luby and Charles Rackoff.
\newblock How to construct pseudorandom permutations from pseudorandom
  functions.
\newblock {\em SIAM Journal on Computing}, 17(2):373--386, 1988.

\bibitem[Mal99]{maldacena1999large}
Juan Maldacena.
\newblock The large-n limit of superconformal field theories and supergravity.
\newblock {\em International journal of theoretical physics}, 38(4):1113--1133,
  1999.

\bibitem[MT88]{morris1988wormholes}
Michael~S Morris and Kip~S Thorne.
\newblock Wormholes in spacetime and their use for interstellar travel: A tool
  for teaching general relativity.
\newblock {\em American Journal of Physics}, 56(5):395--412, 1988.

\bibitem[NDGD06]{nielsen2006quantum}
Michael~A Nielsen, Mark~R Dowling, Mile Gu, and Andrew~C Doherty.
\newblock Quantum computation as geometry.
\newblock {\em Science}, 311(5764):1133--1135, 2006.

\bibitem[PYHP15]{pastawski2015holographic}
Fernando Pastawski, Beni Yoshida, Daniel Harlow, and John Preskill.
\newblock Holographic quantum error-correcting codes: Toy models for the
  bulk/boundary correspondence.
\newblock {\em Journal of High Energy Physics}, 2015(6):149, 2015.

\bibitem[RT06a]{ryu2006aspects}
Shinsei Ryu and Tadashi Takayanagi.
\newblock Aspects of holographic entanglement entropy.
\newblock {\em Journal of High Energy Physics}, 2006(08):045, 2006.

\bibitem[RT06b]{ryu2006holographic}
Shinsei Ryu and Tadashi Takayanagi.
\newblock Holographic derivation of entanglement entropy from the anti--de
  {S}itter space/conformal field theory correspondence.
\newblock {\em Physical review letters}, 96(18):181602, 2006.

\bibitem[RY17]{roberts2017chaos}
Daniel~A Roberts and Beni Yoshida.
\newblock Chaos and complexity by design.
\newblock {\em Journal of High Energy Physics}, 2017(4):121, 2017.

\bibitem[SK79]{Samra_1979}
N~El Samra and R~C King.
\newblock Dimensions of irreducible representations of the classical lie
  groups.
\newblock {\em Journal of Physics A: Mathematical and General},
  12(12):2317--2328, 1979.

\bibitem[SS08]{sekino2008fast}
Yasuhiro Sekino and Leonard Susskind.
\newblock Fast scramblers.
\newblock {\em Journal of High Energy Physics}, 2008(10):065, 2008.

\bibitem[SS14a]{shenker2014black}
Stephen~H Shenker and Douglas Stanford.
\newblock Black holes and the butterfly effect.
\newblock {\em Journal of High Energy Physics}, 2014(3):67, 2014.

\bibitem[SS14b]{shenker2014multiple}
Stephen~H Shenker and Douglas Stanford.
\newblock Multiple shocks.
\newblock {\em Journal of High Energy Physics}, 2014(12):46, 2014.

\bibitem[SS14c]{stanford2014complexity}
Douglas Stanford and Leonard Susskind.
\newblock Complexity and shock wave geometries.
\newblock {\em Physical Review D}, 90(12):126007, 2014.

\bibitem[STU93]{susskind1993stretched}
Leonard Susskind, Larus Thorlacius, and John Uglum.
\newblock The stretched horizon and black hole complementarity.
\newblock {\em Physical Review D}, 48(8):3743, 1993.

\bibitem[Sus14a]{susskind2014addendum}
Leonard Susskind.
\newblock Addendum to computational complexity and black hole horizons.
\newblock {\em arXiv:1403.5695}, 3 2014.

\bibitem[Sus14b]{susskind2014original}
Leonard Susskind.
\newblock Computational complexity and black hole horizons.
\newblock {\em arXiv:1402.5674}, 2 2014.

\bibitem[Sus16]{susskind2016computational}
Leonard Susskind.
\newblock Computational complexity and black hole horizons.
\newblock {\em Fortschritte der Physik}, 64(1):24--43, 2016.

\bibitem[Swi12a]{swingle2012constructing}
Brian Swingle.
\newblock Constructing holographic spacetimes using entanglement
  renormalization.
\newblock {\em arXiv:1209.3304}, 2012.

\bibitem[Swi12b]{swingle2012entanglement}
Brian Swingle.
\newblock Entanglement renormalization and holography.
\newblock {\em Physical Review D}, 86(6):065007, 2012.

\bibitem[Yao82]{yao1982theory}
Andrew~C Yao.
\newblock Theory and application of trapdoor functions.
\newblock In {\em 23rd Annual Symposium on Foundations of Computer Science
  (SFCS 1982)}, pages 80--91. IEEE, 1982.

\bibitem[Yue14]{yuenrandomfunction}
Henry Yuen.
\newblock A quantum lower bound for distinguishing random functions from random
  permutations.
\newblock {\em Quantum Info. Comput.}, 14(13-14):1089--1097, October 2014.

\bibitem[Zha12]{zhandry2012construct}
Mark Zhandry.
\newblock How to construct quantum random functions.
\newblock In {\em 2012 IEEE 53rd Annual Symposium on Foundations of Computer
  Science}, pages 679--687. IEEE, 2012.

\end{thebibliography}

\newcommand{\etalchar}[1]{$^{#1}$}

\appendix

\section{Omitted proof: near-orthogonality of random quantum states}
\label{app:orthogonality}

In this section we present a proof of the fact that

\[\mathbb{E}_U \left| \bra{0^n} (U^\dagger)^K X_1 U^K \ket{0^n}   \right|^2 \leq O\left(\frac{k!}{2^n}\right)\]

To see this, we explicitly write out this expectation as a large sum:

\begin{align}
\label{eq:integral}
&\mathbb{E}_U \left| \bra{0^n} (U^\dagger)^K X_1 U^K \ket{0^n}   \right|^2 =\mathbb{E}_U\displaystyle\sum_{a,b,i_1\ldots i_{k-1}, j_1 \ldots j_{k-1}, i'_1 \ldots i'_{k-1}, j'_1 \ldots j'_{k-1} \in \{0,1\}^n} \\ &(U^\dagger)_{0,i_1}(U^\dagger)_{i_1 i_2}\ldots (U^\dagger)_{i_{k-1} a} U_{\bar{a} j_1}U_{j_1 j_2} \ldots U_{j_{k-1, 0}} (U^\dagger)_{0i'_1} (U^\dagger)_{i'_1 i'_2} \ldots (U^\dagger)_{i'_{k-1} b} U_{\bar{b}j'_1}U_{j'_1 j'_2}\ldots U_{j'_{k-1}, 0}\end{align}
where here the strings $\bar{a},\bar{b}$ denotes the string obtained by flipping the first bit of $a,b$, respectively.

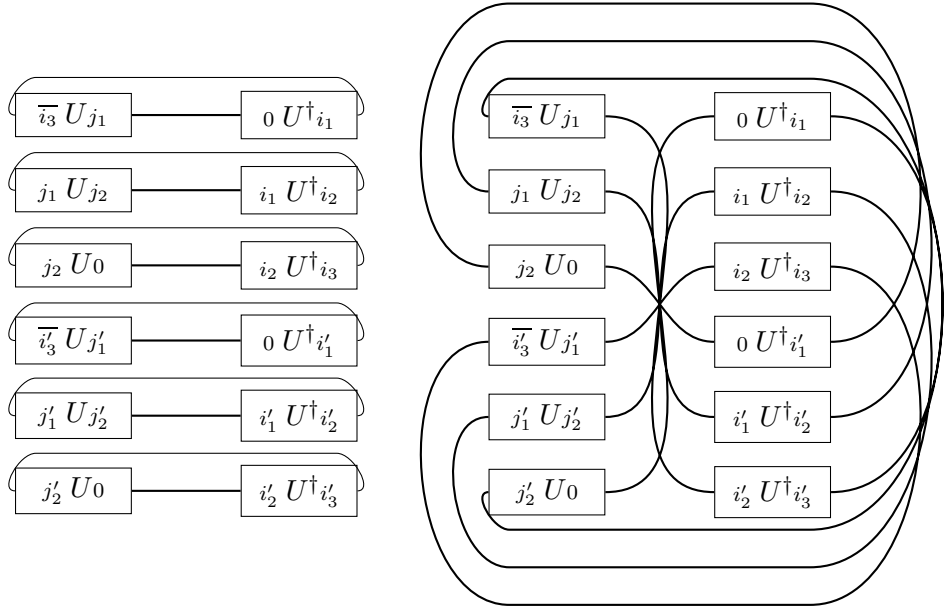
\begin{figure}[h]
    \centering
    
    \begin{tikzpicture}

    \node[rectangle,minimum width=4em,draw] (a0) at (0,0) {\scriptsize $\bar{i_3}$ \normalsize $U$\scriptsize$j_1$};
    \node[rectangle,minimum width=4em,draw] (a1) at (0,-1) {\scriptsize $j_1$ \normalsize $U$\scriptsize$j_2$};
    \node[rectangle,minimum width=4em,draw] (a2) at (0,-2) {\scriptsize $j_2$ \normalsize$U$\scriptsize$0$};
    \node[rectangle,minimum width=4em,draw] (b0) at (0,-3) {\scriptsize $\bar{i'_3}$ \normalsize $U$\scriptsize$j'_1$};
    \node[rectangle,minimum width=4em,draw] (b1) at (0,-4) {\scriptsize $j'_1$ \normalsize $U$\scriptsize$j'_2$};
    \node[rectangle,minimum width=4em,draw] (b2) at (0,-5) {\scriptsize $j'_2$ \normalsize$U$\scriptsize$0$};

    \node[rectangle,minimum width=4em,draw] (c0) at (3,0) {\scriptsize $0$ \normalsize$U^\dagger$\scriptsize$i_1$};
    \node[rectangle,minimum width=4em,draw] (c1) at (3,-1) {\scriptsize $i_1$ \normalsize$U^\dagger$\scriptsize$i_2$};
    \node[rectangle,minimum width=4em,draw] (c2) at (3,-2) {\scriptsize $i_2$ \normalsize$U^\dagger$\scriptsize$i_3$};
    \node[rectangle,minimum width=4em,draw] (d0) at (3,-3) {\scriptsize $0$ \normalsize$U^\dagger$\scriptsize$i'_1$};
    \node[rectangle,minimum width=4em,draw] (d1) at (3,-4) {\scriptsize $i'_1$ \normalsize$U^\dagger$\scriptsize$i'_2$};
    \node[rectangle,minimum width=4em,draw] (d2) at (3,-5) {\scriptsize $i'_2$ \normalsize$U^\dagger$\scriptsize$i'_3$};

    \node[]() at(3,-6.4){};

    \draw [thick] (a0) -- (c0)
    (a1) -- (c1)
    (a2) -- (c2)
    (b0) -- (d0)
    (b1) -- (d1)
    (b2) --(d2);

    \draw (a0) to[out=180,in=180] (-0.5,0.5) -- (3.5,0.5) to[out=0,in=0] (c0);
    \draw (a1) to[out=180,in=180] (-0.5,-0.5) -- (3.5,-0.5) to[out=0,in=0] (c1);
    \draw (a2) to[out=180,in=180] (-0.5,-1.5) -- (3.5,-1.5) to[out=0,in=0] (c2);
    \draw (b0) to[out=180,in=180] (-0.5,-2.5) -- (3.5,-2.5) to[out=0,in=0] (d0);
    \draw (b1) to[out=180,in=180] (-0.5,-3.5) -- (3.5,-3.5) to[out=0,in=0] (d1);
    \draw (b2) to[out=180,in=180] (-0.5,-4.5) -- (3.5,-4.5) to[out=0,in=0] (d2);

\end{tikzpicture}
\begin{tikzpicture}

     \node[rectangle,minimum width=4em,draw] (a0) at (0,0) {\scriptsize $\bar{i_3}$ \normalsize $U$\scriptsize$j_1$};
    \node[rectangle,minimum width=4em,draw] (a1) at (0,-1) {\scriptsize $j_1$ \normalsize $U$\scriptsize$j_2$};
    \node[rectangle,minimum width=4em,draw] (a2) at (0,-2) {\scriptsize $j_2$ \normalsize$U$\scriptsize$0$};
    \node[rectangle,minimum width=4em,draw] (b0) at (0,-3) {\scriptsize $\bar{i'_3}$ \normalsize $U$\scriptsize$j'_1$};
    \node[rectangle,minimum width=4em,draw] (b1) at (0,-4) {\scriptsize $j'_1$ \normalsize $U$\scriptsize$j'_2$};
    \node[rectangle,minimum width=4em,draw] (b2) at (0,-5) {\scriptsize $j'_2$ \normalsize$U$\scriptsize$0$};

    \node[rectangle,minimum width=4em,draw] (c0) at (3,0) {\scriptsize $0$ \normalsize$U^\dagger$\scriptsize$i_1$};
    \node[rectangle,minimum width=4em,draw] (c1) at (3,-1) {\scriptsize $i_1$ \normalsize$U^\dagger$\scriptsize$i_2$};
    \node[rectangle,minimum width=4em,draw] (c2) at (3,-2) {\scriptsize $i_2$ \normalsize$U^\dagger$\scriptsize$i_3$};
    \node[rectangle,minimum width=4em,draw] (d0) at (3,-3) {\scriptsize $0$ \normalsize$U^\dagger$\scriptsize$i'_1$};
    \node[rectangle,minimum width=4em,draw] (d1) at (3,-4) {\scriptsize $i'_1$ \normalsize$U^\dagger$\scriptsize$i'_2$};
    \node[rectangle,minimum width=4em,draw] (d2) at (3,-5) {\scriptsize $i'_2$ \normalsize$U^\dagger$\scriptsize$i'_3$};

    \draw[thick] (a0) to[out=0, in=180] (d2)
    (a1) to[out=0, in=180] (d1)
    (a2) to[out=0, in=180] (d0)
    (b0) to[out=0, in=180] (c2)
    (b1) to[out=0, in=180] (c1)
    (b2) to[out=0, in=180] (c0)
    (a0) to[out=180, in=180] (-0.5,0.5) -- (3.5,0.5) to[out=0,in=0] (d2)
    (a1) to[out=180, in=180] (-0.5,1) -- (3.5,1) to[out=0,in=0](d1)
    (a2) to[out=180, in=180] (-0.5,1.5) -- (3.5,1.5) to[out=0,in=0] (d0)
    (b0) to[out=180, in=180] (-0.5,-6.5) -- (3.5,-6.5) to[out=0,in=0](c2)
    (b1) to[out=180, in=180] (-0.5,-6.0) -- (3.5,-6.0) to[out=0,in=0] (c1)
    (b2) to[out=180, in=180] (-0.5,-5.5) -- (3.5,-5.5) to[out=0,in=0] (c0);

    \end{tikzpicture}

    \caption{Visual representation of our integral \ref{eq:integral} for $k=3$. One connects the right $U$ indices to the left $U^\dagger$ indices via a permutation $\sigma$, and the left $U$ indices with the right $U^\dagger$ indices with a permutation $\tau$. The value of the term is $Wg(\sigma\tau^{-1},2^n)$ if all connected indices are identical and 0 otherwise. Here we illustrate two cases of $\sigma=\tau=Id$ and $\sigma=\tau=\pi$ for the permutation $\pi$ described above. One can easily check that the term on the left -- $\sigma=\tau=Id$ -- has value $0$ because the identifications of the connected indices produces a contradiction. On the other hand the term on the right has every variable identification repeated twice. For example $j_1=i_2'$ is repeated by both connecting the inner indices of the top left and bottom right node, as well as the outer indices of the second from top left and second from bottom right nodes. One can easily check that as the indices range over $\{0,1\}^n$ that this diagram has the maximum number of nonzero terms (i.e. number of settings of $i_1,i_2,\ldots$ in which all variable identification equalities are met). 
    }
    \label{fig:collinssniady}
\end{figure}

Let us now apply the formulae of Collins as Sniady \cite{collins2006integration}. 
We need to consider all ways $\sigma,\tau$ of matching the indices of the $U$ terms to the indices of the $U^\dagger$ terms.
We can denote the visually as follows (see Fig \ref{fig:collinssniady}): draw all $U$ terms with their indices on the left, and all the $U^\dagger$ terms with their indices on the right. Now connect the right $U$ indices to the left $U^\dagger$ indices via a permutation $\sigma$, and the left $U$ indices with the right $U^\dagger$ indices with a permutation $\tau$. The value of the term is $Wg(\sigma\tau^{-1},2^n)$ if all connected indices are identical and 0 otherwise.

In evaluating the order of contribution of a particular choice of $\sigma,\tau$ to our expectation value, two possibly competing considerations come into play. 
First, in the above we are summing over all settings of $a,b,i_1,i_2,\ldots$.
Naively there are $d^{4k-2}$ terms in the sum.
But one can clearly see that these permutations force to identify different variables in this sum (e.g. in Figure \ref{fig:collinssniady} the permutations $\sigma=\tau=Id$ force us to have $j_1=0,j_2=i_2,....$ etc.). Each of these identifications decreases the number of terms in the above sum by a factor of $d$ (or worse -- if these identifications lead to a contradiction like $a=\bar(a)$)-- so one wishes to minimize the number of distinct identifications of indices to maximize the contribution of a particular setting of $\sigma,\tau$.
Second, Collins and Sniady showed that the asymptotics of the Weingarten function are $Wg(\sigma,\tau)=O(d^{-k -|\sigma|})$ where $|\sigma|$ is the minimum number of transpositions needed to produce $\sigma$.
Therefore terms in which $\sigma$ is far from $\tau$ are suppressed by factors of $d$ as well -- and the largest Weingarten function is when $\sigma=\tau$. 

These two notions have the possibility of competing with one another -- e.g. by setting $\sigma$ different than $\tau$ perhaps one can avoid variable identifications and increase the number of nonzero terms in the sum. Fortunately this does not occur in our particular interval -- the minimum number of variable identifications is achieved by setting $\sigma=\tau=\pi$ in the following way, which is captured on the RHS of Figure \ref{fig:collinssniady}.
\[a\rightarrow \bar{b}, j_1 \rightarrow i'_{k-1},\ldots j_{k-1}\rightarrow i'_1, j'_1 \rightarrow i_{k-1} \ldots j'_{k-1} \rightarrow i_{1}\]

This permutation $\pi$ has the special property that all variable identifications occur exactly twice in the matching, so many of them are redundant. 
One can easily compute that the number of remaining terms in the sum over $a,b,i_1,i_2,\ldots$ is precisely $d^{2k-1}$.
As noted previously the asymptotics of the Weingarten function are $O(d^-k)$.
Therefore as a function of $d=2^n$ the asymptotics of this integral are $O(1/d)$ as claimed.

Therefore to complete this proof, we merely need to examine the asymptotics of the function as it depends on $k$ in the case of $k<<d$, particularly when evaluated on $\sigma=Id_{k}$. The Weingarten function (in the case of $k<d$) is defined as
\[ Wg(\sigma\in S_k, d) = \frac{1}{(k!)^2}\sum_{\lambda \vdash k} \frac{\chi^{\lambda}(Id)^2 \chi^\lambda(\sigma)}{s_{\lambda,d}(1)}  \]
where here $\lambda$ is a partition of $k$ labeling the irreps of $S_k$, $\chi^\lambda(\pi)$ denotes the character of $\pi$ under that irrep, and $s_{\lambda, d}(1)$ is the value of the Schur polynomial which for this particular value is equal to the dimension of the irrep of $U(d)$ corresponding to the partition $\lambda$.
Since the character of the identity in any representation is the dimension of the irrep, we have that
\begin{align*}
    Wg(Id_k\in S_k, d) &= \frac{1}{(k!)^2}\sum_{\lambda \vdash k} \frac{\dim_{S_k}(\lambda)^3}{s_{\lambda,d}(1)} \\
    &\leq \frac{1}{(k!)^2} \max_{\lambda' \vdash k}\frac{ \dim_{S_k}(\lambda')}{s_{\lambda',d}(1)} \sum_{\lambda \vdash k} \dim_{S_k}(\lambda)^2 \\
    &= \frac{1}{k!} \max_{\lambda' \vdash k}\frac{ \dim_{S_k}(\lambda')}{\dim_{U(d)}(\lambda')} 
\end{align*}
where in the last line we used the fact that the sum of dimensions squared of all irreps of a group is the order of the group. Here $\dim_G(\lambda)$ is the dimension of the irrep $\lambda$ of group $G$ -- we overload notation as partitions of $k$ both correspond to irreps of $S_k$ and irreps of $U(d)$.
Therefore to complete the bound we merely need to upper bound this ratio as a function of $k,d$. To do this, for the numerator we make use of the crude upper bound that
\[\dim_{S_k}(\lambda)\leq (k!)^{1/2}\]
since the sums of the dimensions squared of irreps of a group sum it its order. 
For the denominator we make use of the formulae for the dimensions of such irreps. Given a partition of $k$, the dimension of the corresponding irrep of $U(d)$ is given by \cite{Samra_1979}:
\[\dim_{U(d)}(\lambda) = \prod_{ij} \frac{d-i+j}{\lambda_i -j + \tilde{\lambda}_j -i+1} \]
where the $ij$ range over the entries of the Young diagram corresponding to the partition $\lambda \vdash k$, and the $\lambda_i$ ($\tilde{\lambda}_j$) refer to the row (column) hook lengths at index $ij$.
Since $1 \leq i,j \leq k$ at all points in the diagram (and $k<d$ here) we have that
\[\dim_{U(d)}(\lambda)\geq \Omega\left(\frac{(d-k)^k}{(2k)^k}\right) = \Omega(d^k/k^{O(k)})\]

Putting these together with Stirling's approximation we obtain that
\[
Wg(Id_k\in S_k, d) = O\left(\frac{k^{O(k\log k)}}{d^k}\right)
\]
as desired.

\section{Pseudo-complexity}
\label{sec:pseudocomplexity}
Consider the following definition:

\begin{definition}
The \emph{pseudo-complexity} of a quantum circuit $C$ and an initial state $\ket{\phi}$, denoted $\mathcal{PC}_{\epsilon}(C,\ket{\phi})$, is defined as follows: Take the gates of the circuit $C$ from beginning to end, and greedily apply local rewrite rules to shorten the circuit (for instance, by cancelling out gates with their adjacent inverses) so long as the error incurred by each rewrite rule is $\leq\epsilon$ in operator norm.
\end{definition}

In our setting we imagine $C$ is the (Trotterized) circuit arising from the time evolution $e^{-i\hamilcft t}$ (possibly perturbed by operator insertions) and $\ket{\phi} = \ket{TFD}$. Here we consider $\epsilon$ to be a small parameter much less than the length of the circuit, so that the end state $C\ket{\phi}$ produced by the simplified circuit is still close to the original one.
Also by ``apply local rewrite rules'' we mean greedily perform operations given by a finite list of rules on how to simplify circuits. For instance the rewrite rule list could include cancelling out gates with their adjacent inverses, or commuting a Pauli through a Clifford gate and cancelling, etc. 

Clearly pseudo-complexity is manifestly computable in polynomial time -- because it is computable by a greedy algorithm applied to the circuit -- and is robust to small perturbations.
It also exhibits the desired properties of complexity. 
For example, it should grow linearly for an exponential amount of time under local time evolution, because the (Trotterized) circuit of a general local Hamiltonian evolution will not have local cancellations. 
Also, pseudo-complexity should have the same response to local operator insertions as complexity (at least in the general case), in that it should exhibit ``switchback effects'' driven by local cancellations in the circuit.
This was is one of the key pieces of evidence for Complexity=Volume \cite{stanford2014complexity} as the volume has analogous behavior.

Of course pseudo-complexity has its shortfalls.
For one, it is not time symmetric -- given $C,\ket{\phi}, \ket{\psi}=U\ket{\phi}$ we might have that $\mathcal{PC}_{\epsilon}(\ket{\psi},C,\ket{\phi})\neq \mathcal{PC}_{\epsilon}(\ket{\phi},C^{-1},\ket{\psi}) $.
This is because the local rewrite rules are applied greedily -- and a non-greedy version of this definition (such as ``apply these local rewrite rules in the most optimal way possible'') would likely itself be $\mathsf{NP}$-hard to compute. 
For another this measure grows linearly forever and does not saturate at exponential time, as occurs with complexity and would be expected to occur in a theory of quantum gravity in AdS living in a finite-dimensional Hilbert space.
Therefore Pseudo-complexity=Volume would not predict a recurrence in the wormhole volume back to a short wormhole at doubly exponential time, as is predicted by Complexity=Volume.
More fundamentally, pseudo-complexity is not a function of the quantum state of the CFT, but depends on additional information about the CFT state -- namely its ``history'' or how it evolved to its present state.
Therefore positing pseudo-complexity=Volume would require a fundamental modification of the AdS/CFT dictionary map to depend on more than the CFT state alone.

\end{document}